\documentclass[12pt,a4paper,twoside,twocolumn,english,english]{article}
\usepackage{setspace}  %% Zur Setzung des Zeilenabstandes
\usepackage{babel}     %% Sprachen-Unterstuetzung
\usepackage{calc}      %% ermoeglicht Rechnen mit Laengen und Zaehlern
\usepackage[T1]{fontenc}       %% Unterstutzung von Umlauten etc.
\usepackage[utf8]{inputenc}  %% 
\usepackage{ulem}
\usepackage{adjustbox}
\usepackage{graphicx}
\usepackage[font=footnotesize,labelfont=bf]{caption}
\usepackage[list=true, font=footnotesize, labelfont=bf, labelformat=parens, position=top]{subcaption}
\usepackage{here}

\usepackage{listings}
\usepackage{listingsutf8}
\usepackage{appendix}
\usepackage{amsmath,amssymb,amsfonts,amstext} %% zusaetzliche Mathe-Symbole

\usepackage{lmodern} %% type1-taugliche CM-Schrift als Variante zur
\usepackage[comma,numbers,sort&compress,super]{natbib}

\usepackage{booktabs}                      %% Befehle fuer besseres Tabellenlayout
\usepackage{longtable}                     %% umbrechbare Tabellen
\usepackage{array}                         %% zusaetzliche Spaltenoptionen

\usepackage{textcomp,gensymb}

\newcommand{\beginsupplement}{%
        \setcounter{table}{0}
        \renewcommand{\thetable}{S\arabic{table}}%
        \setcounter{figure}{0}
        \renewcommand{\thefigure}{S\arabic{figure}}%
     }

\usepackage{units}   %% Variante fuer Einheiten

\usepackage[pdfstartview=FitH,      % Oeffnen mit fit width
            breaklinks=true,        % Umbrueche in Links, nur bei pdflatex default
            bookmarksopen=true,     % aufgeklappte Bookmarks
            bookmarksnumbered=true  % Kapitelnummerierung in bookmarks
            ]{hyperref}
\usepackage{xcolor}
\hypersetup{
    colorlinks,
    linkcolor={black},
    citecolor={black},
    urlcolor={black}
}

\usepackage{epsf}
\usepackage{authblk} %Add author affiliations

\title{Aluminum depletion induced by complex co-segregation of carbon and boron in a $\Sigma5\,[3\,1\,0]$ bcc-iron grain boundary}
\author[1]{Ahmadian, A.\thanks{a.ahmadian@mpie.de}}
\author[2]{Scheiber, D.}
\author[1]{Zhou, X.}
\author[1]{Gault, B.}
\author[1]{Liebscher, C. H.}
\author[2]{Romaner, L.}
\author[1]{Dehm, G.}
\affil[1]{Max-Planck-Institut fuer Eisenforschung GmbH, 40237 Düsseldorf, Germany}
\affil[2]{Materials Center Leoben GmbH, 8700 Leoben, Austria}
\date{}                     %% if you don't need date to appear
\begin{document}

\maketitle

\begin{otherlanguage}{english}
\begin{abstract}
The local variation of grain boundary atomic structure and chemistry caused by segregation of impurities influences the macroscopic properties of poylcrystalline materials. Here, the effect of co-segregation of carbon and boron on the depletion of aluminum at a $\Sigma 5\,(3\,1\,0\,) [0\,0\,1]$ tilt grain boundary in a $\alpha-$~Fe-$4~at.~\%$Al bicrystal was studied by combining atomic resolution scanning transmission electron microscopy, atom probe tomography and density functional theory calculations. The atomic grain boundary structural units mostly resemble kite-type motifs and the structure appears disrupted by atomic scale defects. Atom probe tomography reveals that carbon and boron impurities are co-segregating to the grain boundary reaching levels of >1.5 at.\%, whereas aluminum is locally depleted by 
approx. 2~at.\%. First-principles calculations indicate that carbon and boron exhibit the strongest segregation tendency and their repulsive interaction with aluminum promotes its depletion from the grain boundary. It is also predicted that substitutional segregation of boron atoms may contribute to local distortions of the kite-type structural units. These results suggest that the co-segregation and interaction of interstitial impurities with substitutional solutes strongly influences grain boundary composition and with this the properties of the interface.   
\end{abstract}

\section*{Introduction}
Segregation of solutes and impurity atoms to grain boundaries (GBs) have been studied extensively in the last decades because of their strong influence on the macroscopic physical and mechanical properties of materials \cite{papazian_grain_1971-1, suzuki_effect_1983, hansel_grain_1986, lejcek_temperature_1990, lejcek_segregation_1991, lejcek_characterization_1994, fraczkiewicz_boron_1998, fraczkiewicz_influence_2000, krakauer_atomic_1993, maruyama_interaction_2003}. Typically, GB segregation is studied in binary alloys serving as model systems to discern segregation tendencies of individual solutes \cite{papazian_grain_1971-1, lejcek_temperature_1990, krakauer_atomic_1993, maruyama_interaction_2003}. However, technologically relevant alloy systems adopt a complex composition, which is tailored to maximize their mechanical properties. The interplay of different solutes in ternary or higher order systems on grain boundary segregation is far less understood \cite{seah_grain_1980, xing_solute_2018}. Xing et al. \cite{xing_solute_2018} developed a simple thermodynamic model to capture the competing mechanisms in ternary alloys that decrease, increase or do not affect GB segregation. However, this model has limited applicability when non-metallic elements are involved, which play a critical role in understanding embrittlement effects in iron-based alloys or steels \cite{hansel_grain_1986, guttmann_equilibrium_1975}.        

Grain boundary segregation in iron (Fe) is often characterized by Auger electron spectroscopy (AES) providing excellent chemical sensitivity, but limited spatial resolution leaving the structure of the GB elusive. However, it could be shown that sulphur (S) and phosphorous (P) are segregating to GBs \cite{grabke_surface_1986, grabke_equilibrium_1977} in $\alpha-$~Fe promoting embrittlement of the material. To prevent this detrimental tendency, carbon (C) or boron (B) can be introduced, which hinder the diffusion of S and P to GBs and therefore enhance their cohesive strength \cite{tauber_grain_1978, pichard_influence_1976, liu_effect_1992}.

The isolated role of B on increasing GB cohesion has long been anticipated in a range of material systems \cite{liu_effect_1985, liu_effect_1992, painter_effects_1987, wu_first_1994} and it has been shown that the addition of small amounts of B partially suppresses intergranular fracture in brittle intermetallic FeAl alloys \cite{lejcek_boron_2003, fraczkiewicz_boron_1998, fraczkiewicz_influence_2000}. It is generally also assumed that C acts as a GB cohesion enhancer and recent ab-initio calculations found that the theoretical fracture strength of $\sum$5 tilt GBs increases with increasing C excess concentration \cite{wang_first-principles_2016}. 

Aluminum (Al) contained in small quantities in ferritic alloys as a substitutional alloying element was also found to eliminate intergranular brittleness \cite{rellick_elimination_1970}. It was proposed that it hinders the segregation of residual oxygen (O) to GBs upon quenching by forming Al-oxide particles. Rellick et al. \cite{rellick_elimination_1970} suggested that the interaction of C with O will mitigate embrittlement emphasizing the critical role of solute interactions. However, Yuasa et~al. \cite{yuasa_first-principles_2013} concluded from density functional theory (DFT) calculations that Al grain boundary segregation will reduce the ductility of Fe due to a change in electronic structure from metallic to covalent-like bonding type. In contrast, Geng et~al. \cite{geng_influence_2001} showed through first-principles calculations that Al may acts as a weak cohesion enhancer in agreement with the Rice-Wang model \cite{rice_embrittlement_1989}. These conflicting results indicate that it is of great interest to investigate the GB segregation behavior of Al in Fe and to explore the impact of prominent impurity elements such as B and C on segregation tendencies. 

Deciphering co-segregation phenomena requires a complete understanding of the atomic grain boundary structure and composition as well as description of the interaction between impurities \cite{guttmann_equilibrium_1975,lejcek_grain_2010}. The correlation of atom probe tomography (APT) and transmission electron microscopy (TEM) has paved the way to obtain novel insights into the segregation behavior of solutes even in GBs of complex alloys \cite{raabe_grain_2014}. However, the underlying atomistic effects and the interplay of elements on the grain boundary segregation behavior are often uncharted.

%This was observed in DFT simulations of the Fe-C-Mn system \cite{wicaksono_interaction_2017}, where the C segregation enhanced the enrichment of Mn at the GB. Their results were consistent with the APT work of van Landeghem et~al. \cite{van_landeghem_solute_2016}, who observed a high Mn content at the GB in the Fe-C-Mn but none in the Fe-Mn-N system. 

The discrete atomic GB structure and the presence of defects at the GB are additional factors impacting the segregation behavior. It is known that C and B tend to segregate to sites of high stress fields such as dislocation cores \cite{da_rosa_co-segregation_2017}.  
Therefore it is indispensable to resolve the GB structure and composition at atomic resolution to relate the underlying segregation mechanisms by ab-initio calculations. For example, Medlin et~al. \cite{medlin_defect_2017} investigated an asymmetric $\Sigma 5$ GB in pure Fe using scanning transmission electron microscopy (STEM). They found that the GB adopts a complex topography by dissociating into nanofacets consisting of $(2\,1\,0)$ and $(3\,1\,0)$ facets, which are separated by GB facet junctions. However, connecting such complex GB structure and the segregation of minor impurities like S, P, C and B, which are even relevant in pure Fe, is extremely challenging by STEM alone. Besides using AES to study GB segregation in ferritic Fe and steel \cite{lejcek_temperature_1990, liu_effect_1992, lejeek_anisotropy_nodate-3}, APT \cite{krakauer_atomic_1993, maruyama_interaction_2003} is capable of providing 3D compositional information with high elemental sensitivity. Aberrations arising from the field evaporation process \cite{Larson2013a, Ashton2020} limit the spatial resolution of APT effectively to approx. 0.5nm \cite{DeGeuser2020} for precipitates. For  grain boundary structure and other crystalline defects, the spatial resolution is expected to be in a similar range or better \cite{Jenkins2020}. Ultimately, the combination of atomic resolution STEM and APT provides structural as well as compositional information of the GB with the highest spatial and elemental resolution \cite{Zhou2016, liebscher_strain-induced_2018-1, liebscher_tetragonal_2018, makineni_diffusive_2018, stoffers_correlating_2017, Herbig2018}.

In the present work, we correlate atomic resolution STEM, energy-dispersive X-ray spectroscopy (EDS) and APT measurements on a $\Sigma 5\,(3\,1\,0)[0\,0\,1]$ tilt GB in $\alpha-$~Fe-$4~at.\%$Al to resolve the atomic GB structure and its local composition. The GB predominantly adopts a kite-type structural unit and the presence of GB defects is indicated. The Al concentration at the GB is observed to decrease, whereas the impurities B and C are clearly segregating. Complementary first-principles based DFT calculations are employed to elucidate the segregation tendencies of Al, B and C and shed light onto the impact of solute interactions on the co-segregation behavior. A kinetic model is employed to investigate the temperature and time dependent segregation of solutes for similar conditions as in the experiment.

\section*{Results}
\subsection*{Experimental results}
\textbf{Atomic GB structure.}\\ A low-magnification high angle annular dark-field (HAADF)-STEM image of the $\Sigma 5\,(3\,1\,0)$ GB observed in $[0\,0\,1]$ direction is shown in Fig.~\ref{haadf_adf_overview}~a). The local contrast in Fig.~\ref{haadf_adf_overview}~a) is mainly attributed to changes in sample thickness and atomic mass, where at a constant sample thickness the intensity is proportional to the atomic number $Z^{1.5-2}$ \cite{pennycook_scanning_2011-1}. At this length scale the GB appears straight and the formation of GB defects, such as steps or facets can not be resolved. The dark contrast of the GB may be attributed to GB grooving, a lower atomic density or lattice strain, causing dechannelling of the electron beam. Figure~\ref{haadf_adf_overview}b) shows the simultaneously acquired low-angle ADF-STEM image revealing bright contrast along the GB, which may be attributed to the accumulation of strain fields at the GB \cite{liebscher_strain-induced_2018-1}. The red square in Fig.~\ref{haadf_adf_overview}~a) indicates a segment of the boundary, which was imaged at atomic resolution with a local sample thickness of about $40\,nm$ as determined by electron energy loss spectroscopy (EELS) measurements.

\begin{figure}[ht]
  \centering
\includegraphics[width=0.98\linewidth]{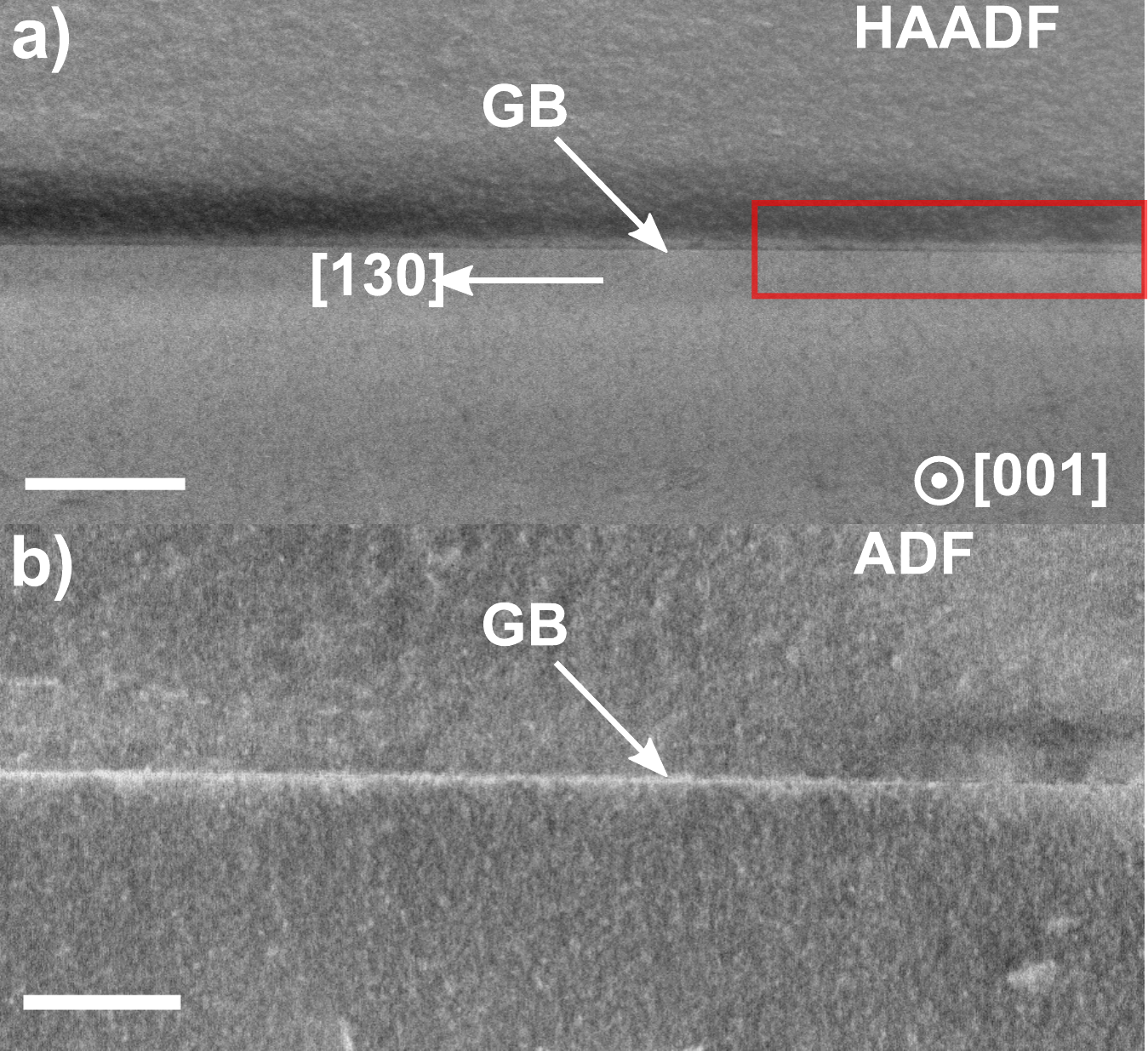}
\caption{\textbf{Overview images of the $\Sigma 5\,(3\,1\,0\,) [0\,0\,1]$ GB} (a) HAADF-STEM image and (b) corresponding low-angle ADF-STEM image showing bright contrast at the GB core indicating an accumulation of strain fields at the GB. The red rectangle in (a) indicates regions, which were imaged at atomic resolution. The scale bar is $100\,nm$.}
\label{haadf_adf_overview}
\end{figure}

A representative atomic resolution HAADF-STEM image of the GB (Fig.~\ref{HAADF_FFT})) shows the projected structural units of the $\Sigma 5\,(3\,1\,0)[0\,0\,1]$ GB viewed along the $[0\,0\,1]$ direction. The blue and red squares in the corresponding fast Fourier transform (FFT) shown in the inset of Fig.~\ref{HAADF_FFT}~a) highlight the $<$110$>$ type reflections of the lower and upper grains, respectively. Both grains are misoriented by $\sim$38$^\circ$ around the common $[0\,0\,1]$ tilt axis. This observation is in good agreement with the EBSD analysis presented in supplementary Figure~\ref{supp:EBSD}. A Butterworth Fourier-filter was applied to the image to remove low-frequency noise and highlight the atom columns at the boundary as shown in Fig.~\ref{HAADF_FFT}~b). A magnified view reveals the kite-type structural units of the $\Sigma 5\,(3\,1\,0)[0\,0\,1]$ GB, which is consistent with the predicted structure of a pure bcc-Fe GB \cite{cak_first-principles_2008, medlin_defect_2017}. However, it should be noted, that a weak signal is apparent in the open regions of the kites (indicated by yellow arrows). Besides, the kite-structure of the GB is distorted on the right-hand side in Fig.\ref{HAADF_FFT}~a) (indicated by a red dashed rectangle). These features and their origins will be discussed later.

\begin{figure}[htbp]
\includegraphics[width=0.98\linewidth]{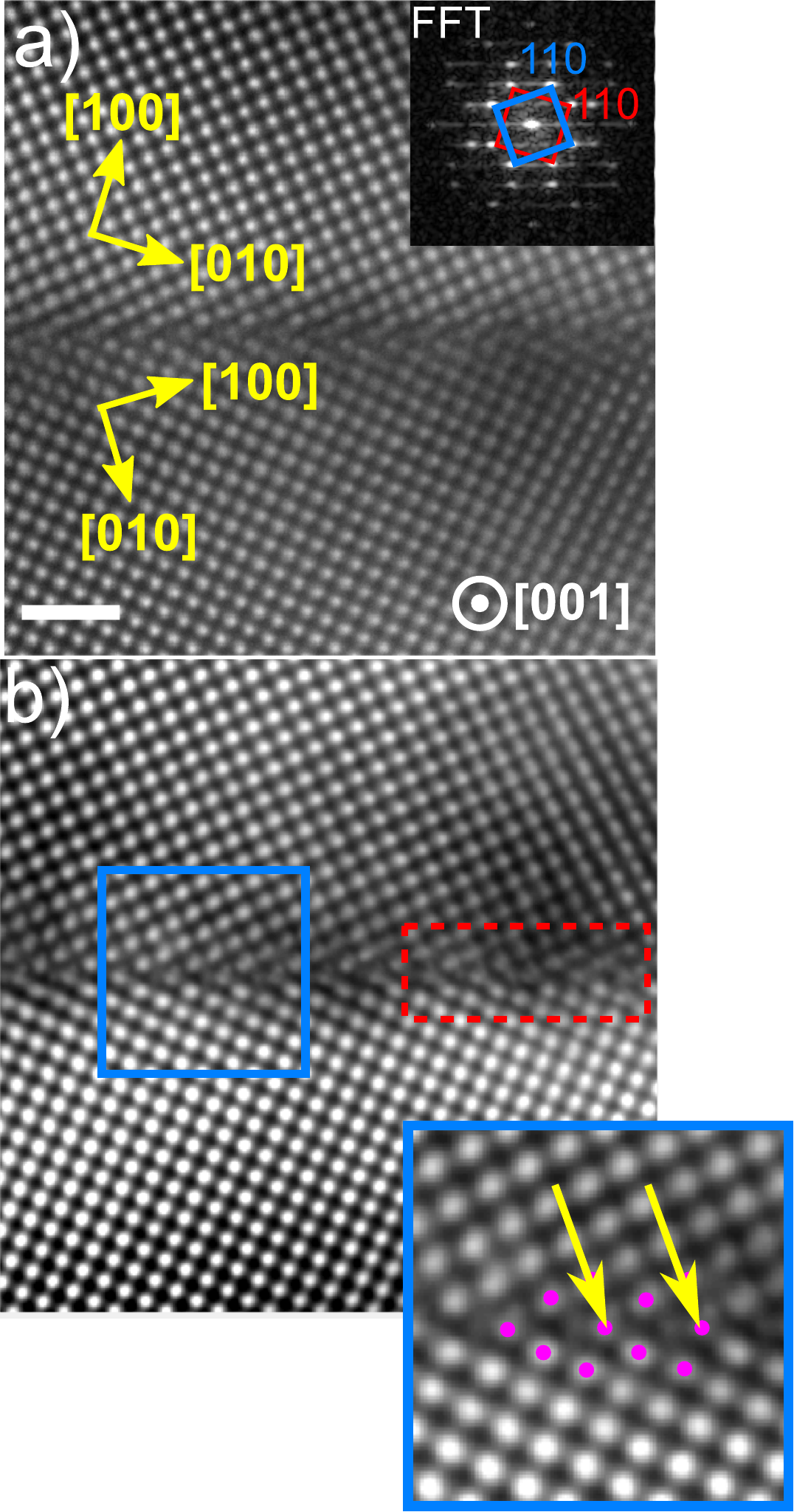}
\caption{\textbf{Atomic structure of the $\Sigma 5\,(3\,1\,0\,) [0\,0\,1]$ GB in the Fe-4at.$\%$Al bi-crystal.} (a) Raw HAADF-STEM image and corresponding FFT. From the FFT the misorientation angle is determined to $38^\circ$. (b) Fourier-filtered image of (a). The inset shows a magnified view of the kite-type structural GB units highlighted by magenta dots. The yellow arrows indicate positions where weak contrast in the kite centers is observed. The scale bar is $1\,nm$.}
\label{HAADF_FFT}
\end{figure} 

Besides kite-type structural units, defects such as steps (see supplementary Figure~\ref{supp:HAADF}) or facets are present in other locations of the GB, which are introduced due to a local change in GB inclination.  Figure~\ref{facetted}~a) shows a HAADF-STEM image of a GB segment intersecting an Al$_{2}$S$_{3}$ precipitate. The local GB curvature is increased in close vicinity of the precipitate as indicated by a black rectangle in Figure~\ref{facetted}~a). Atomic resolution imaging reveals a nano-facetted GB (Fig.~ref{facetted}~b). It can be seen that while the lower grain is in $[0\,0\,1]$ orientation, the upper grain is slightly off zone axis. The inclination angle of the GB with respect to the $(3\,1\,0)$ plane is $\approx 7^\circ$ and the GB facets are adopting the symmetric $\Sigma 5\,(3\,1\,0)$ kite structure (blue box in Figure~\ref{facetted}~b)). The magnified view of a step (red box in Figure~\ref{facetted}~b)) reveals that it resembles a single kite structural unit, which is rotated by $\sim87^\circ$ with a height of $\sim0.46\,nm$. It should be mentioned that in the investigated TEM specimen only one Al$_{2}$S$_{3}$ precipitate (see supplementary Figure~\ref{supp:EDS}) was found, but even though a rather large GB curvature was introduced, the symmetric GB segments exhibit the kite-type structure.\\

\begin{figure}[htbp]
\includegraphics[width=0.98\linewidth]{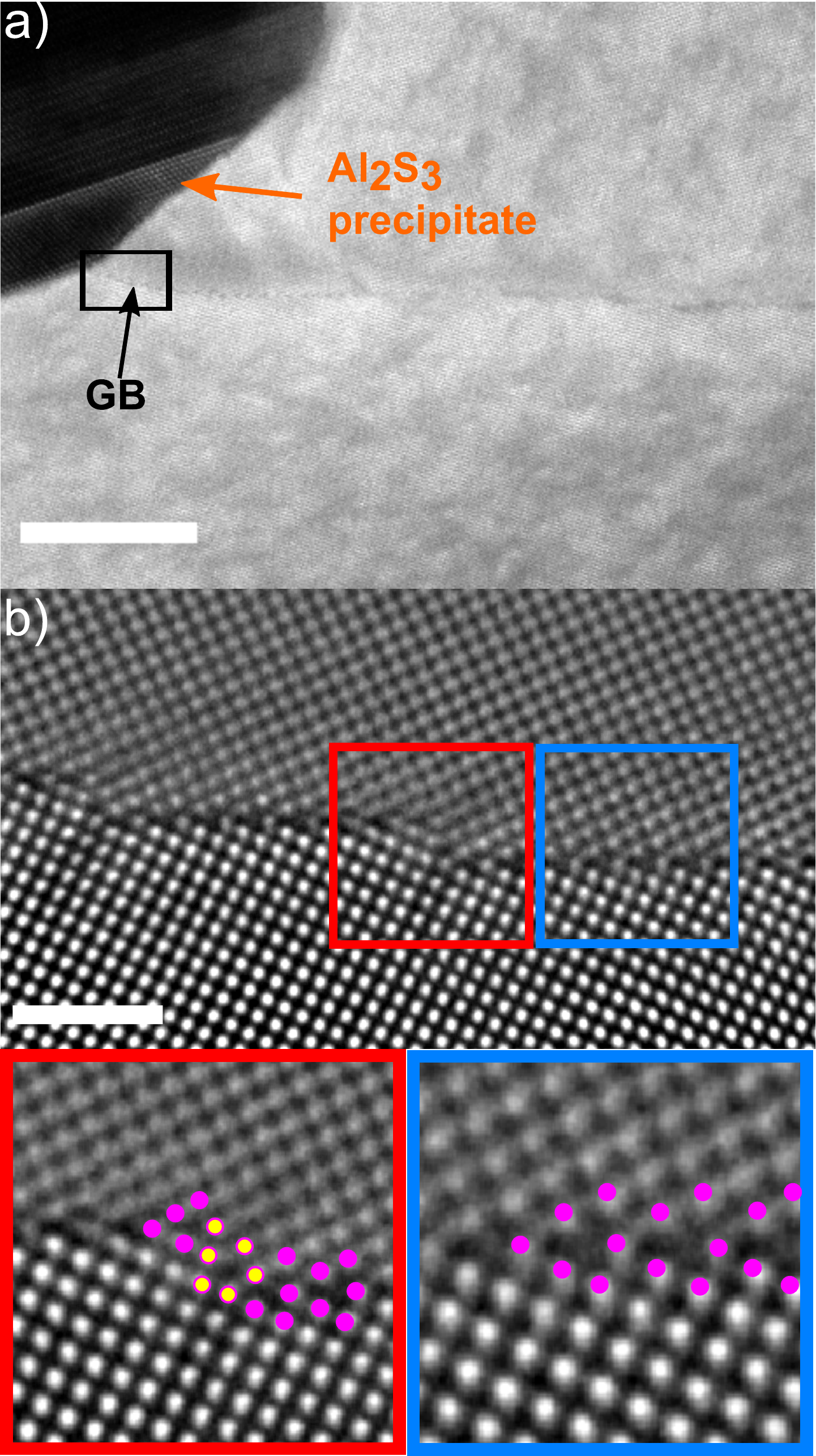}
\caption{\textbf{Grain boundary nano-facetting in close vicinity to a GB precipitate}~a) HAADF-STEM image taken near the Al$_2$S$_3$ precipitate. The black rectangle indicates region where the boundary is bend. b) Atomic resolution image of the region within the black rectangle showing nm-sized facets. The blue and red boxes indicate the symmetric facet and the step, respectively, as well as magnified views of the corresponding structural units. The scale bar is $20\,nm$ in a) and  $1\,nm$ in b)}.
\label{facetted}
\end{figure}

\noindent
\textbf{Grain boundary segregation.}\\
In parts of the GB, the atomic arrangement has shown slight deviations form the perfect kite structure and the presence of a GB precipitates indicates that impurity segregation could impact the GB structure. Besides $\sim4\,at\%$ Al, the as-grown bicrystal contains different impurities such as C, B, P, S, and Si in the lower ppm range. STEM-EDS elemental mapping was used to determine the GB composition as shown in Figure~\ref{EDX}~a). Here, the Al-K elemental map is superimposed on the simultaneously acquired HAADF-STEM image, where the dotted line highlights the position of the GB and the blue and red rectangles indicate regions where EDS spectra were integrated from. No Al enrichment at the GB is visible from the EDS map. To obtain a qualitative understanding of Al segregation, the spatial difference method was applied where the spectrum from within the grains (blue) is subtracted from the GB spectrum (red). Each integration window had an area of $3~\times~10~\unit{nm^2}$ in a region of constant thickness of $40\,nm$ as proved by STEM-EELS. The two spectra of the grain interior were averaged and for comparison all spectra were normalized according to the Fe-K$_{\alpha}$ edge as illustrated in Fig.~\ref{EDX}~b). The difference spectrum (orange curve) has a non-zero value at the Fe-K$_\alpha$ edge, which is not caused by a slight shape difference of the peaks within the grain and at the GB. Therefore, all peaks  in the spectra were fitted by Gaussian peaks (solid lines). A magnified view of the Al-K$_\alpha$ peak is shown in Figure~\ref{EDX}~c). A decrease in Al intensity of $\sim$5$\times10^{-3}$ or 12\% with respect to the grain interior is observed at the GB. The standard deviation of the noise level of the difference spectrum around the Al-K$_\alpha$ peak is $\sim$1.3$\times10^{-3}$, which is nearly $\times$5 smaller than the peak maximum. This analysis suggests that Al is depleted at the GB, although first-principles calculations predicted a tendency for Al to segregate to substitutional sites at an idealized $\Sigma 5\,(3\,1\,0)$ GB in bcc-Fe (position $3$ in Fig.~\ref{fig:gbseg}). Any sign of enrichment of other impurity elements were not revealed in the STEM-EDS data. Since measuring light elements is difficult in STEM-EDS, APT was used in addition to get a precise insight of the GB chemistry.\\

\begin{figure*}[htbp]
  \includegraphics[width=0.98\linewidth]{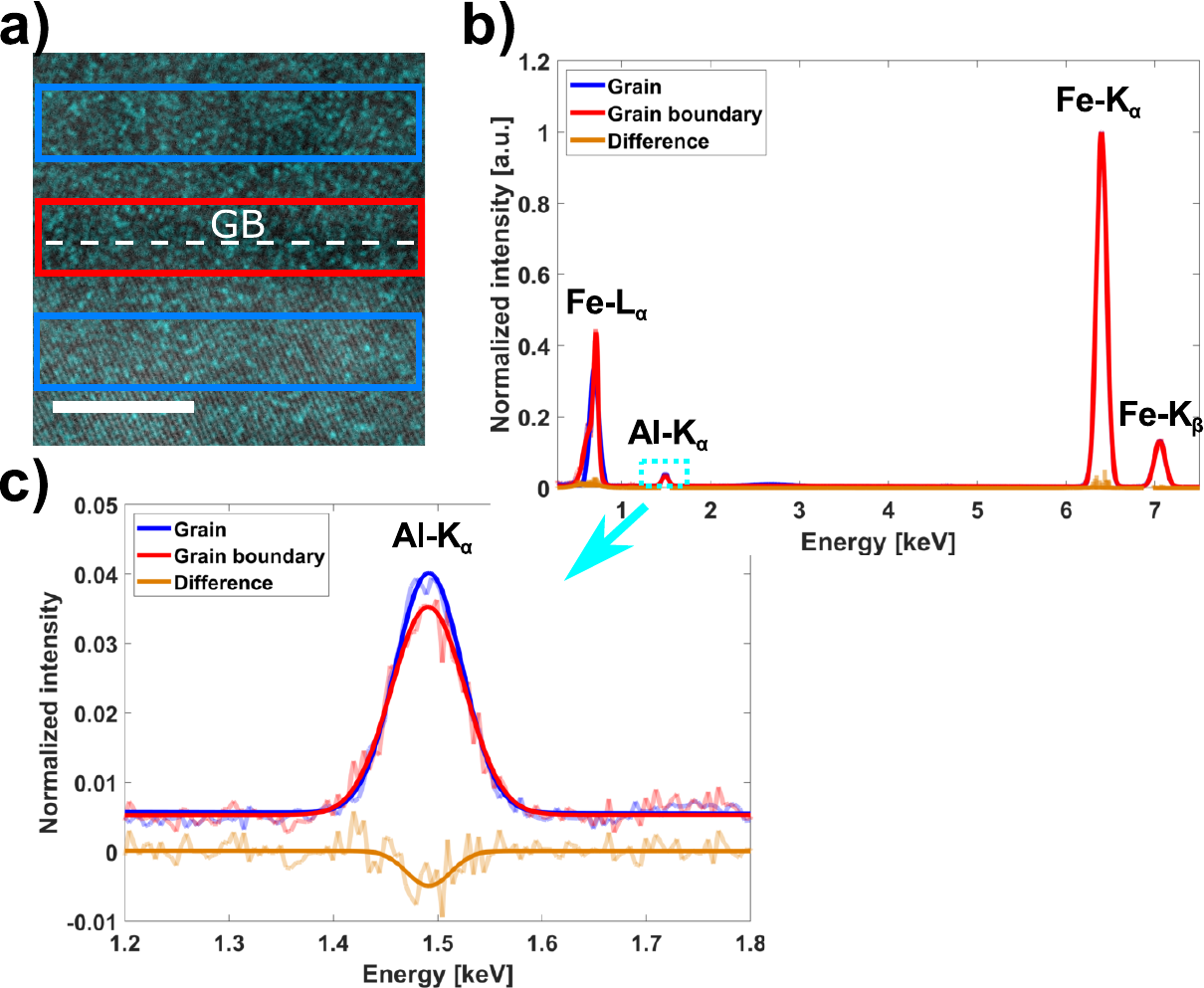}
\caption{\textbf{Elemental distribution of Al at the $\Sigma 5\,(3\,1\,0)$ GB.} a) Overlay of the HAADF-STEM image and the Al-K elemental map taken across the $\Sigma 5\,(3\,1\,0)$ GB. The colored boxes indicate the location where the spectra in b) were summed. b) Summed EDS spectra, which were normalized to the Fe-K$_{\alpha}$ peak, extracted from the grain interior (blue) and the GB (red). The corresponding difference spectrum is shown in orange. In order to minimize effects from differences in peak shapes, the spectra were fitted by a Gaussian function. c) A magnified view of the Al-K$_\alpha$ edge indicates a slight decrease of Al at the GB. The semitransparent curves in b) and c) display the raw spectra, the solid lines the Gaussian fit. The Scale bar in a) is $5\,nm$.}
\label{EDX}
\end{figure*}
 
\noindent
\textbf{Grain boundary composition.}\\ To obtain insights in the local distribution of Al and other impurity elements such as C and B at the GB with highest possible elemental sensitivity, we performed correlative STEM-APT investigations of the GB. Figure~\ref{correlative}~a) presents a low magnification HAADF-STEM image of the needle shaped APT specimen containing the $\Sigma 5$ GB at a distance of $\sim100\,nm$ from the apex. From the width of the needle-shaped specimen, the thickness of the tip at the boundary region can be estimated to $\sim70\,nm$. The viewing direction is along the $[0\,0\,1]$ tilt axis. Figure~\ref{correlative}~b) shows the distribution of Al, C and B atoms. In contrast to C and B enrichment at the GB, a depletion of Al is apparent. A linear 1D composition profile extracted from a cylindrical region with $20\,nm$ in diameter (see Figure~\ref{correlative}~b)) positioned perpendicular to the GB is shown in Figure~\ref{correlative}~c). The Al concentration decreases from $\approx 3.8\,at.\%$ in the bulk (which is in agreement to our wet chemical analysis) to $\approx 2.4\,at.\%$ at the GB. In contrast, the C and B concentration increase to $2.4\,at.\%$ and $1.8\,at.\%$, respectively. A more profound analysis provides information about the Gibbs interfacial excess (IE) $\Gamma_i$ of solute i, which describes the number of atoms per unit area at interfaces \cite{krakauer_atomic_1993}. The measurements show that $\Gamma_B \approx 2.4\,\text{atoms}/nm^2$ and $\Gamma_C \approx 3.1\,\text{atoms}/nm^2$, where $\approx 8\,\text{atoms}/nm^2$ correspond to one monolayer. It should be noted that the IE value could increase when the diameter of the cylinder was reduced and placed across a solute enriched region within the GB, indicating that the elemental distribution is inhomogeneous. For Al a negative IE value of $\Gamma_{Al} \approx -3.1\,\text{atoms}/nm^2$ was obtained.   
 
 \begin{figure*}[htbp]
  \includegraphics[width=0.98\linewidth]{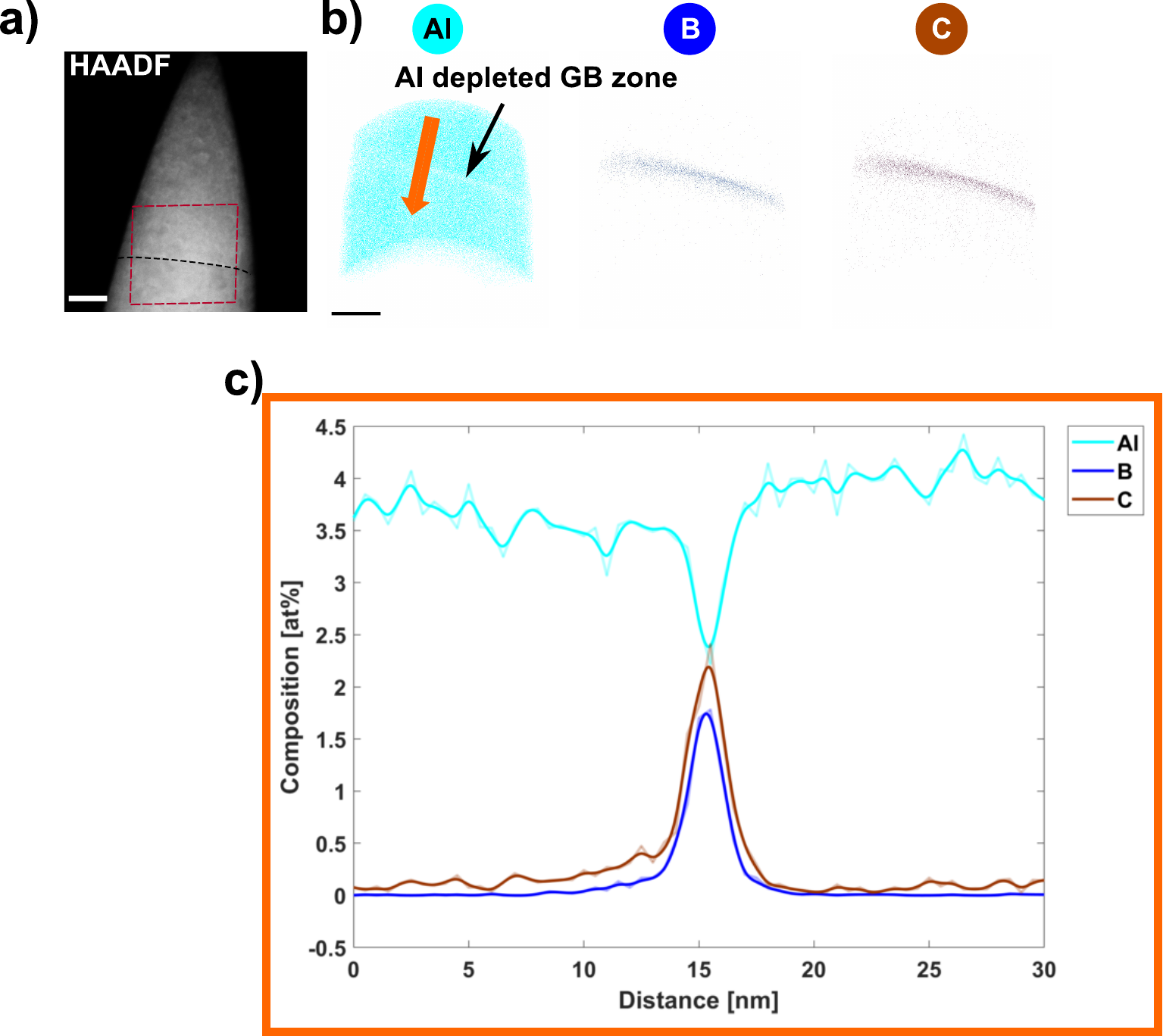}
\caption{\textbf{Correlative STEM-APT study of the $\Sigma$\,5 GB.} a) HAADF-STEM image of the needle shaped APT specimen with both grains in [0\,0\,1] orientation. The GB is labelled with a black dashed line. b) The atom maps of Al, B and C reconstructed from the volume outlined by a red rectangle in a). At the GB, a depletion of Al and enrichment of B and C is observed. c) The corresponding composition profiles extracted from a cylindrical region marked by an orange arrow in b) across the GB shows a decrease of Al and increase of B and C concentration at the GB. Scale bar in a) is $20\,nm$ and in b) $10\,nm$}
\label{correlative}
\end{figure*}
 
For better understanding of C and B distribution, a second APT specimen was extracted from a location $\sim 6\,mm $ apart from the previous STEM-APT observations. Figure~\ref{APT}~a) shows the reconstructed $3D$ volume of the needle shaped specimen, where this time the GB plane was chosen to be perpendicular to the evaporation direction ($[0\,0\,1]$ points towards the evaporation direction). Again, a clear enrichment of C (brown) and B (blue) at the GB were observed, but no segregation of Al. Rotation  of the reconstruction by $90^\circ$ (see Fig.~\ref{APT}~b)) is showing the solute distribution along the GB plane, i.e. viewing along the $[3\,1\,0]$ direction. The distribution of B and C within the GB plane seems to be different with C showing a rather periodic segregation pattern with C rich segments every $25\,nm$. The composition profile along the GB (green arrow in a)) is shown in Fig.~\ref{APT}~d) confirming a difference in the segregation behaviour between C and B. A modulating distribution of C with an amplitude of $\sim 1~at.\%$ and a periodicity of $25\,nm$ was obtained. On the other hand, the average content of B is $\sim 1 \,at.\%$ with no significant modulation.
 
 Although Al exhibits a high solubility of $4\%$ at room temperature in bcc-Fe \cite{kubaschewski_ironbinary_2013} a clear depletion at the GB was observed, which is in contradiction to previous computational \cite{yuasa_first-principles_2013, geng_influence_2001, scheiber2020solute} and experimental \cite{lejcek_solute_2004, mintz_hot-ductility_1979-1} studies. The complex co-segregation of C and B revealed by APT in this study indicates that those interstitial elements might strongly affect the segregation behavior of Al.  
 
\begin{figure*}[htbp]
  \includegraphics[width=0.98\linewidth]{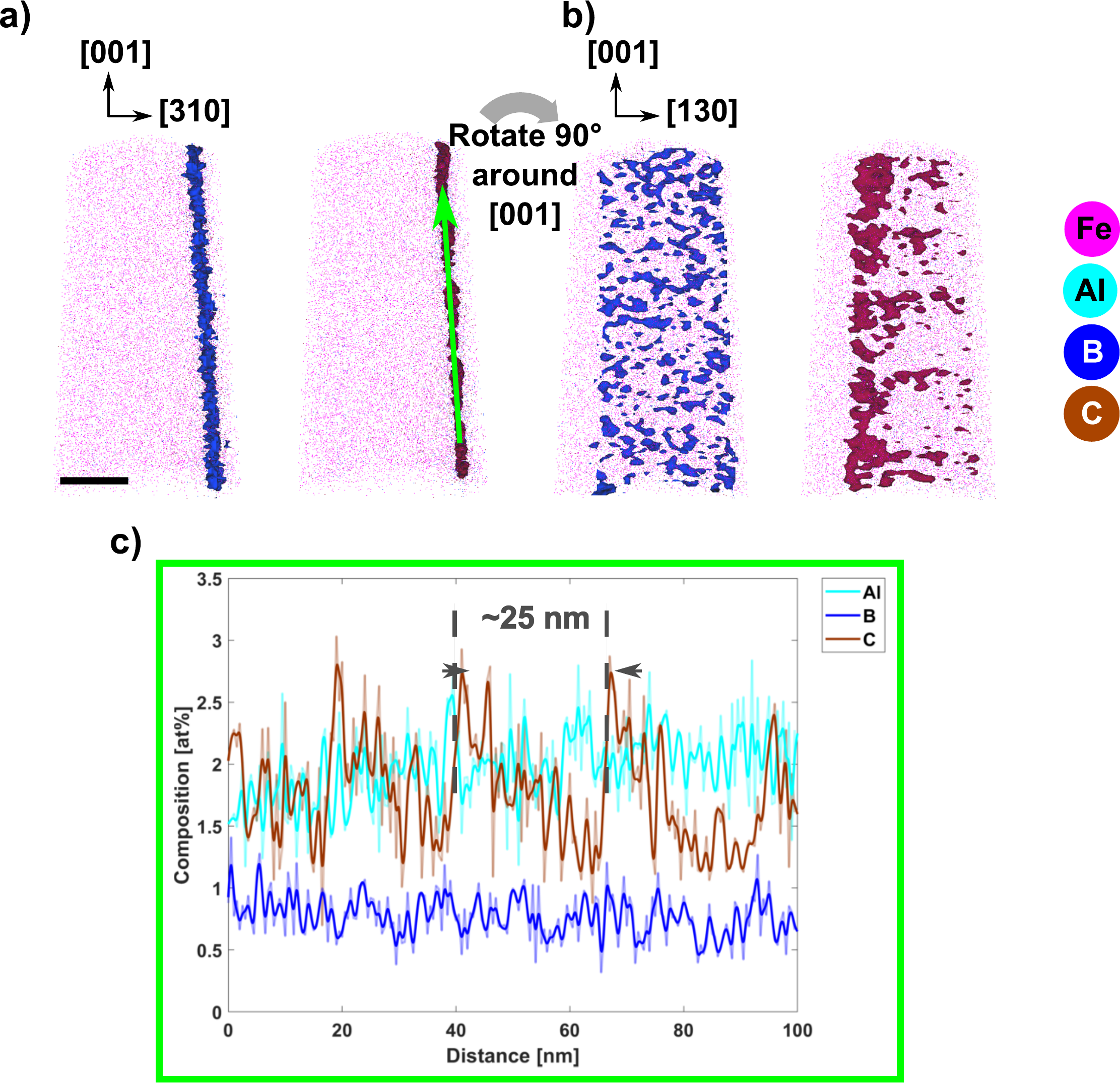}
\caption{\textbf{Depletion of Al and co-segregation of C and B at the $\Sigma 5\,(3\,1\,0)$ GB.} a) 3D APT reconstruction showing C (brown) and B (blue) segregation for isoconcentration surfaces of $1 \,at.\%$ viewed perpendicular to the GB plane. b) The same APT reconstruction rotated by $90^\circ$ around $[0\,0\,1]$ such that the viewing direction is along the $(3\,1\,0)$ GB plane. c) Corresponding composition profile of Al across the GB extracted in a cylindrical volume with $5\,nm$ diameter along the orange arrow in a). d) Corresponding composition profile of the GB extracted in a cylinder with $5\,nm$ diameter along the green arrow in a). While the B concentration is nearly constant at $\sim 1\,at. \%$ along the boundary, the C concentration varies periodically. The average distance between the regions with a peak C concentration of $\sim$2.3~at.$\%$ is determined to $25\,nm$. Scale bar in a) is $20\,nm$}
\label{APT}
\end{figure*}

\subsection*{Computational results}
We performed DFT calculations to explore the interaction of solutes at the $\Sigma 5\,(3\,1\,0)$ GB and their effect on segregation energies. A particular focus was laid on the interaction of C and B with Al to elucidate the observed Al depletion. Furthermore, the segregation kinetics are investigated to obtain insights into the complex co-segregation during bicrystal fabrication.

The pristine $\Sigma 5\,(3\,1\,0)[0\,0\,1]$ GB was modelled with 78 atoms in the unit cell and the ground state structure was determined using the $\gamma$-surface approach \cite{scheiber_ab_2016} at 0K. The resulting structure viewed along the [0\,0\,1] tilt axis is shown in Fig.~\ref{fig:gbseg}.a. In a first step we have calculated the segregation energies of Al, C, and B at the GB using the equation

\begin{equation}
E_{seg,i} = \left(E_{gb,i}^x - E_{gb}\right) - \left(E_{bulk}^x - E_{bulk}\right),
    \label{eq:eseg}
\end{equation}

where $E_{gb}$ is the total energy of the pristine GB and $E_{gb,i}^x$ denotes the total energy of the GB cell with the solute $x$ at GB site $i$. The site $i$ is varied over positions shown in the top panel of Fig. \ref{fig:gbseg}. The reference of the solute in the bulk is given by the energy difference of $E_{bulk}$, i.e. a cubic bulk cell that contains 128 Fe atoms, and $E_{bulk}^x$, which corresponds to the total energy of the bulk cell with solute $x$ at its preferred position. The position in the bulk is substitutional for Al, while for C and B it is the octahedral interstitial site \cite{fors_nature_2008}. Because the atomic volume of B is in between typical substitutional and interstitial solutes, we consider B as both interstitial and substitutional at the GB. For the computation of the segregation energy, the number of atoms needs to be kept constant, which is the case in Eq. \ref{eq:eseg} if the site type is the same in the bulk and at the GB, i.e. either interstitial or substitutional. For placing B in a substitutional site at the GB and in an interstitial site in the bulk, the segregation energy needs to be compensated by the energy of a single Fe atom in the bulk $E_{Fe}=E_{bulk}/128$:

\begin{equation}
E_{seg,i} = \left(E_{gb,i}^x - E_{gb}\right) - \left(E_{bulk}^x - E_{bulk}\right) + E_{Fe}.
    \label{eq:esegsubB}
\end{equation}

A negative segregation energy of a solute indicates that the solute favours segregation to the GB.

%\begin{figure}[htbp]
%\includegraphics[width=0.98\linewidth]{s5_cell.png}
%\caption{Structure of the pristine $\Sigma 5\,(0\,1\,3)[1\,0\,0]$ GB showing three unit cells where the green lines mark the periodic boundaries of a single unit cell.}
%\label{fig:gbpure}
%\end{figure}

\begin{figure}[htbp]
%\begin{subfigure}[c]{1.0\linewidth}
\includegraphics[width=1\linewidth]{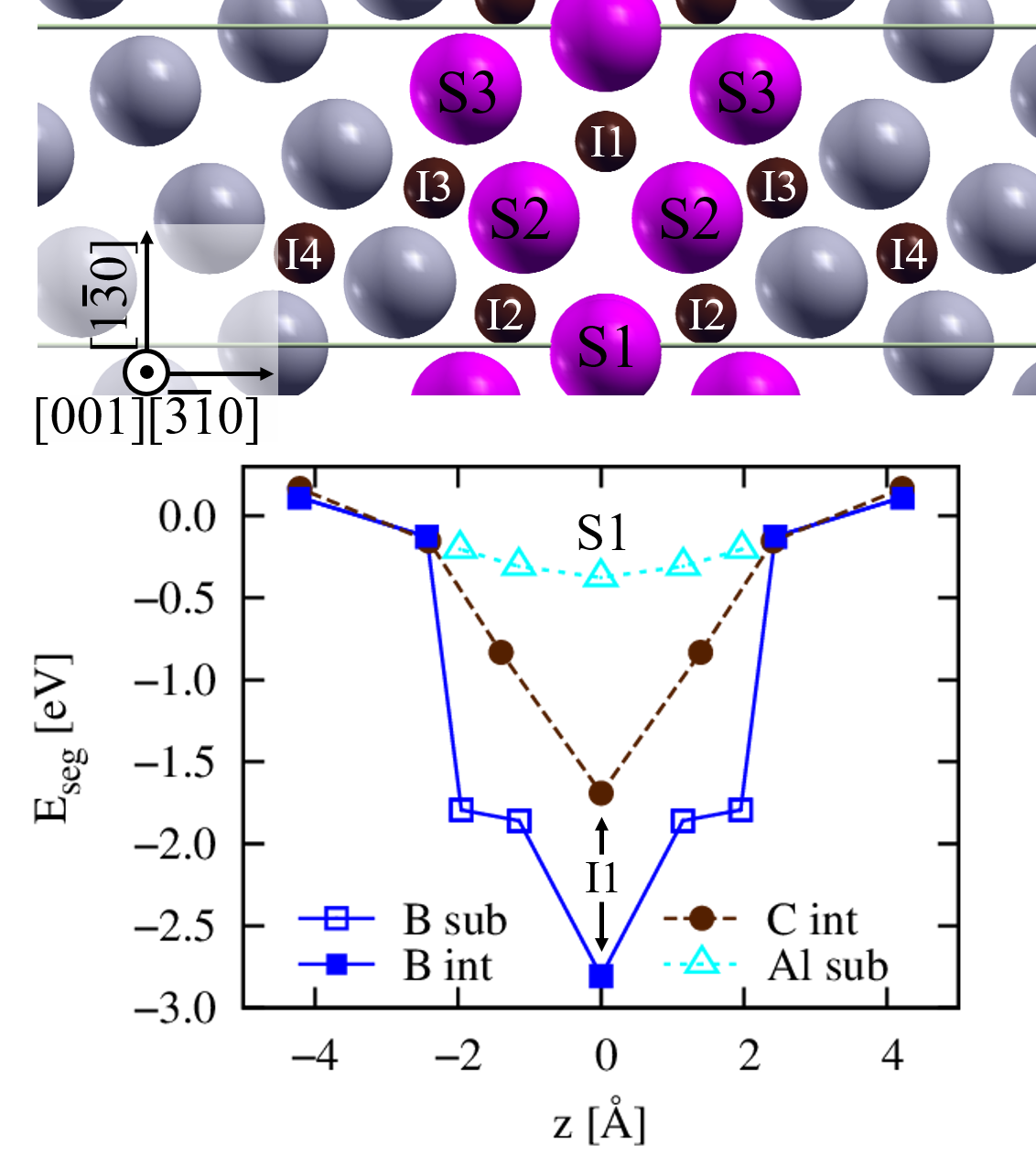}
%\caption{}
%\end{subfigure}
%\begin{subfigure}[r]{0.9\linewidth}
%\includegraphics[width=1\linewidth]{EbindVsDist.png}
%\caption{}
%\end{subfigure}
\caption{\textbf{Solute segregation energies at the GB.} The top panel 
%of a) 
shows the GB 
%(rotated by 90 counter-clockwise with respect to Fig. \ref{fig:gbpure}) 
with the interstitial sites (I1-I4), where C and B were placed, and substitutional sites (S1-S3). The bottom panel 
%of a) 
contains a plot of segregation energies of Al to substitutional sites (empty symbols), C to interstitial sites (solid symbols), and B to both substitutional and interstitial sites. 
%In b), the binding energies of Al with B and with C are shown as a function of the distance between the solutes.
}
\label{fig:gbseg}
\end{figure}

We computed segregation energies at a coverage of 0.5 monolayers (ML) for Al to sites S1-S3, for C to sites I1-I4 and for B to all labelled sites in the top panel of Fig. \ref{fig:gbseg}. The resulting segregation energies  for Al, B, and C are given in Fig. \ref{fig:gbseg}~a). By far the strongest segregation tendency is observed for B with a segregation energy of $-2.8\,\unit{eV}$ to the interstitial site I1 at the GB center. The second strongest segregation tendency is found for the substitutional site S2 next to the GB center with $-1.86\,\unit{eV}$. For C and Al the strongest segregation tendency is observed to the GB center with $-1.7\,\unit{eV}$ in the case of C for position I1 and only $-0.38\,\unit{eV}$ for Al for position S1, respectively. The segregation tendency decreases with increasing distance from the GB. When placing B at the segregation site I2, the B atom moved to site I1.

\begin{figure*}[htbp]
\centering
\begin{subfigure}[c]{0.24\linewidth}
\includegraphics[width=1\linewidth]{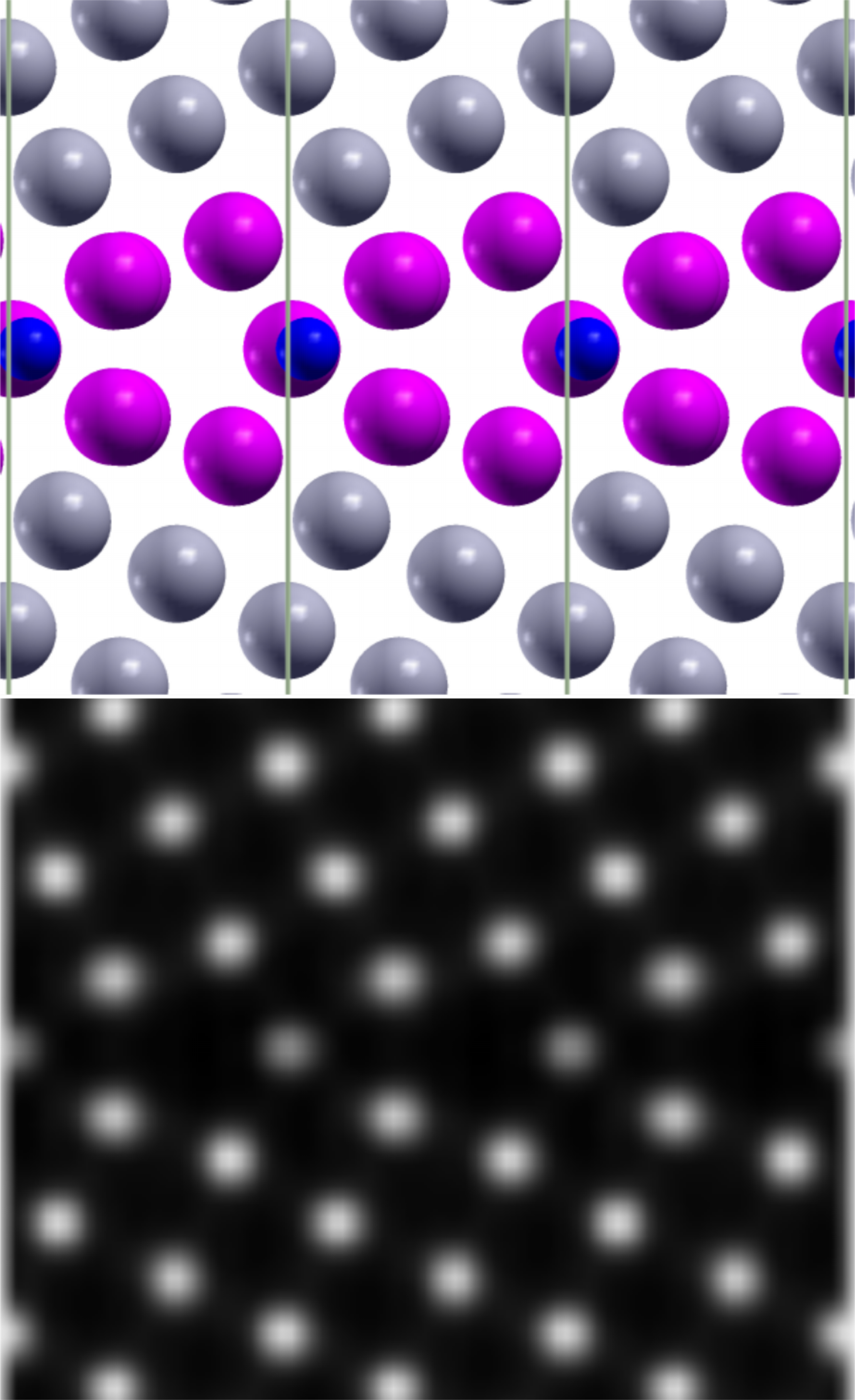}
\caption{}
\end{subfigure}
\begin{subfigure}[c]{0.24\linewidth}
\includegraphics[width=1\linewidth]{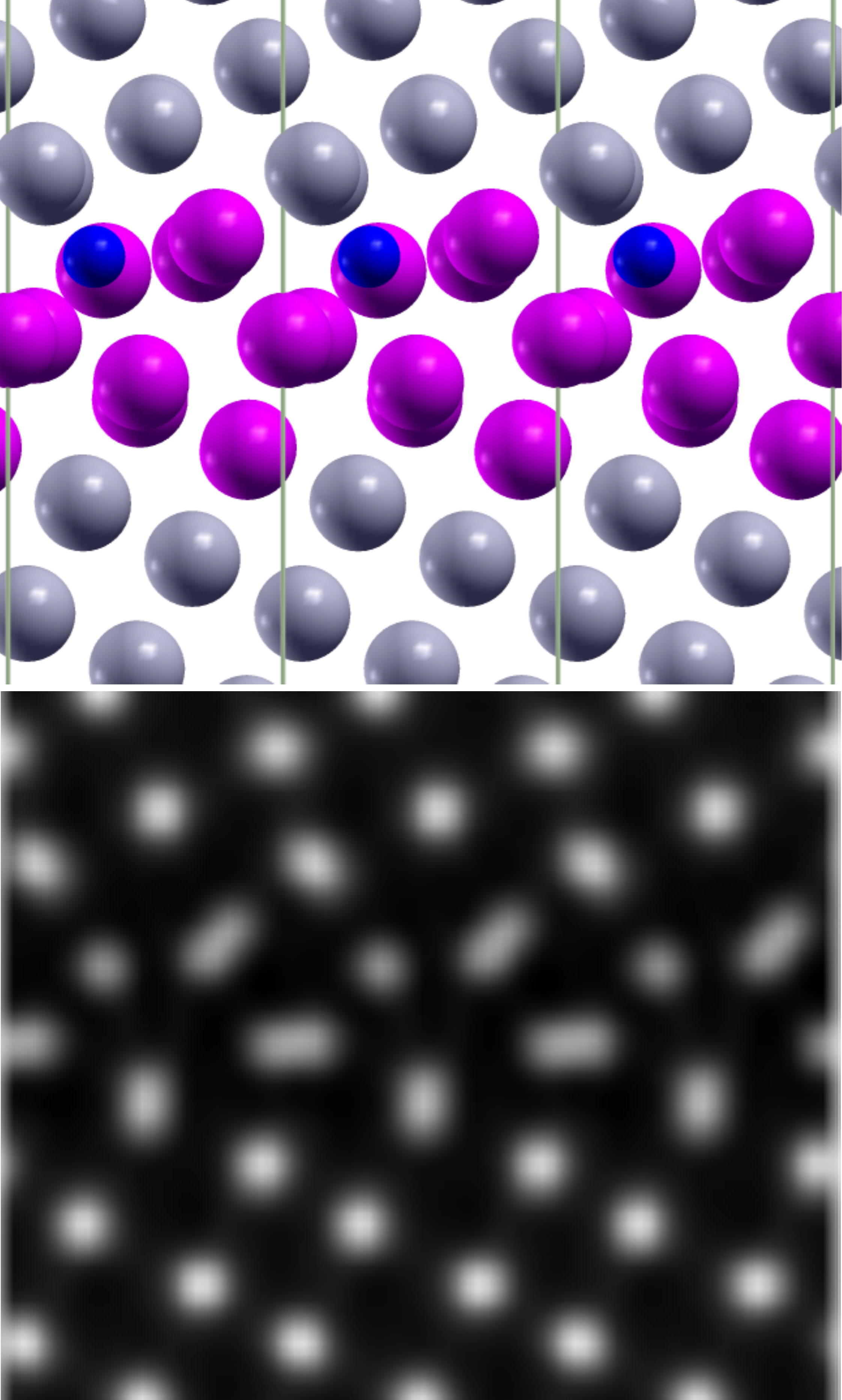}
\caption{}
\end{subfigure}
\begin{subfigure}[c]{0.24\linewidth}
\includegraphics[width=1\linewidth]{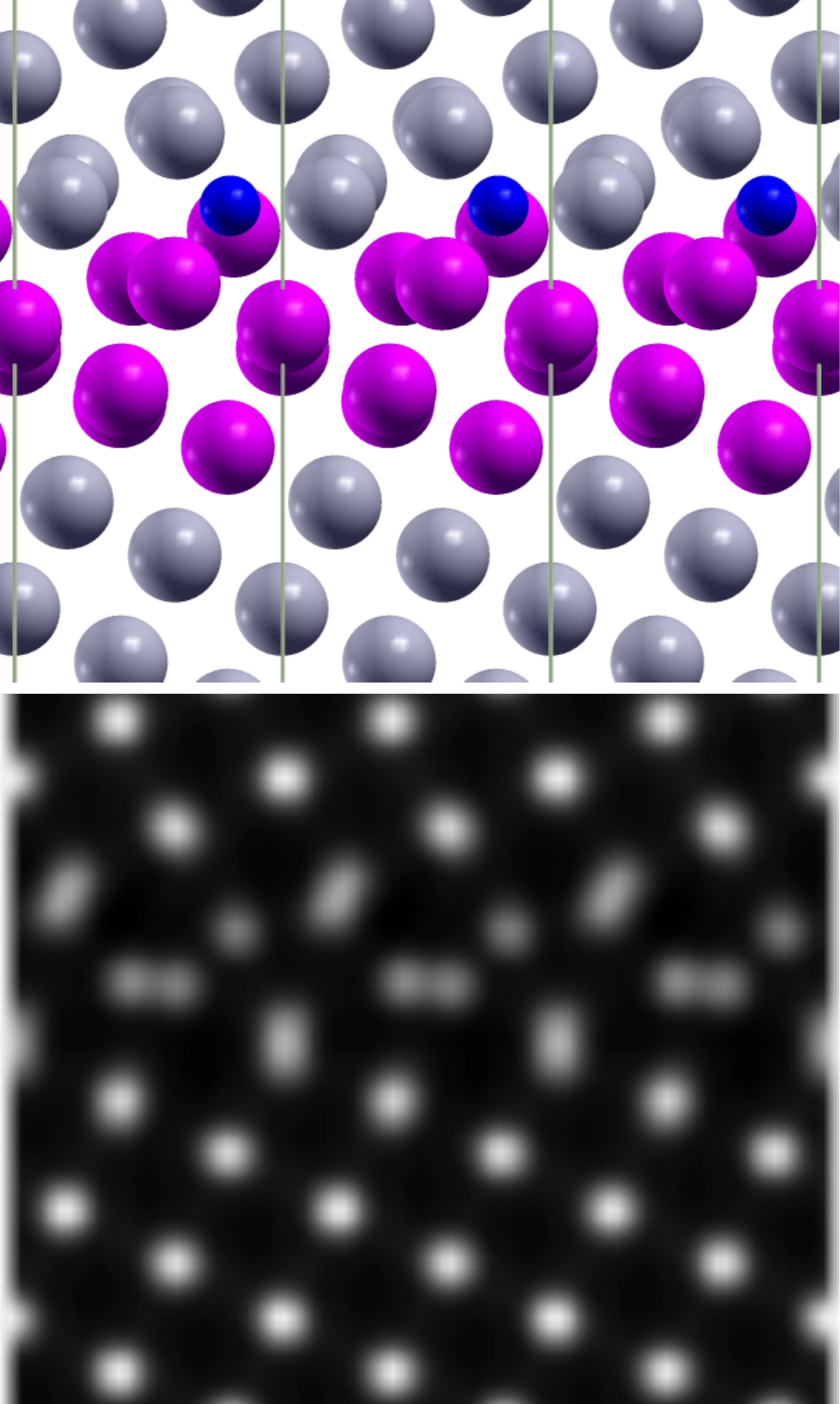}
\caption{}
\end{subfigure}
%\begin{subfigure}[c]{0.24\linewidth}
%\includegraphics[width=1\linewidth]{s5_B_s2_m.png}
%\caption{}
%\end{subfigure}
\begin{subfigure}[c]{0.24\linewidth}
\includegraphics[width=1\linewidth]{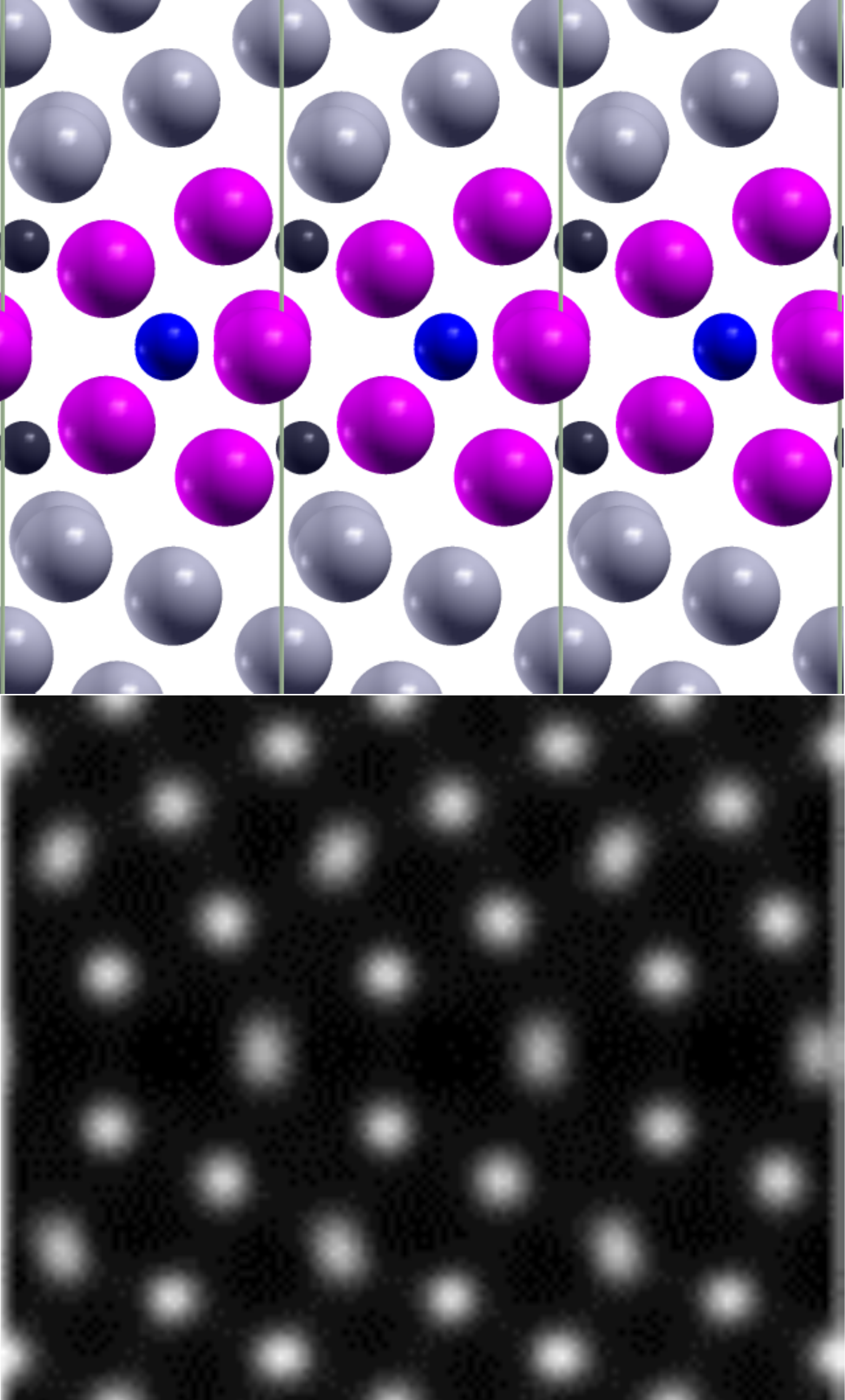}
\caption{}
\end{subfigure}
\caption{Disorder at the GB for B at substitutional segregation sites. The upper panels of a), b) and c) show B at substitutional sites 1, 2, 3. The lower panels are the corresponding HAADF-STEM simulated structures. Panel d) shows the GB segregated with two B (in one column) and two C atoms from Fig.~\ref{fig:gbsegcov}~b.
%Panel d) show B at site 3 with different GB structure highlighting, respectively.
}
\label{fig:Bsub}
\end{figure*}

The segregation tendency of B to substitutional sites S2 and S3 at the GB is substantial, but it also introduces a strong displacement of the surrounding Fe atoms and hence a distortion of the kite structure as seen in Fig. \ref{fig:Bsub}. Especially for B at site S2, the B induced distortion is significant. This can be quantified by the mean squared displacement (MSD) of the atoms in the vicinity of B in a radius of 5 \AA{}. For B at site S1, S2, and S3 the MSD is 0.14, 0.40, and 0.59 \AA{}, respectively. The corresponding simulated HAADF-STEM images using the Prismatic software \cite{ophus_fast_2017, pryor_streaming_2017} are shown in the lower panels of Fig. \ref{fig:Bsub}. Here, the simulated HAADF contrast predominately shows the Fe atomic columns and the B and/or C atoms are only indirectly visible. This can be seen in the slight decrease in intensity of the atomic columns in the kite tips when B is segregating to the substitutional site S1 (see Fig.~\ref{fig:Bsub}~a)).  When B is segregating to position S2 (Fig.~\ref{fig:Bsub}~b)), the neighboring Fe atoms are displaced causing a smearing of the HAADF contrast and the individual columns are no longer resolvable. An even stronger displacement of the Fe atoms is introduced when B is located at site S3, as seen in the top panel of Fig.~\ref{fig:Bsub}~c), and here Fe dumbbells are observed at the kite tips in the simulated HAADF-STEM image (lower panel of Fig.~\ref{fig:Bsub}~c)). Even placing B as well as C in the intersitial positions I1 and I2 (Fig.~\ref{fig:Bsub}~d)) can have a slight distortion of the kite tips, where the Fe atoms are shifted perpendicular to the boundary plane.      
%Due to this distortion it seems as if the center of the GB has moved by one layer, which is shown in Fig. \ref{fig:Bsub} panels c) and d) by a different highlighting of the GB motif.

So far, the computed segregation energies are obtained for individual solutes, but the interaction of solutes can have significant effects on the segregation tendencies when the binding energies are in the range of the segregation energies \cite{aksyonov2016impact,scheiber2018impact}. In the following section, we explore the co-segregation behavior of Al with both C and B as a possible explanation for the observed depletion of Al from the GB. To that end, we have computed the binding energies of Al at its preferred site S1 in the GB center to B and C at vicinal interstitial segregation sites (I1-I4). The binding energy between Al and a solute $x$ is defined as

\begin{equation}
E_{bind}^{Al-x} = \left(E_{Al} + E_{x}\right) - \left(E_{Al-x} + E\right),
    \label{eq:ebind}
\end{equation}

where $E_{Al-x}$ denotes the total energy of the GB cell with Al and the solute $x$ present. $E_{Al}$ and $E_{x}$ are the energies of the same cell with either Al or solute $x$ placed at their lowest energy segregation sites, respectively. $E$ is the energy of the simulation cell not containing any solutes. For attractive interactions, the binding energy is positive, while repulsive interactions are associated with negative binding energies. 

\begin{figure}[htbp]
\begin{subfigure}[c]{1.0\linewidth}
\includegraphics[width=1\linewidth]{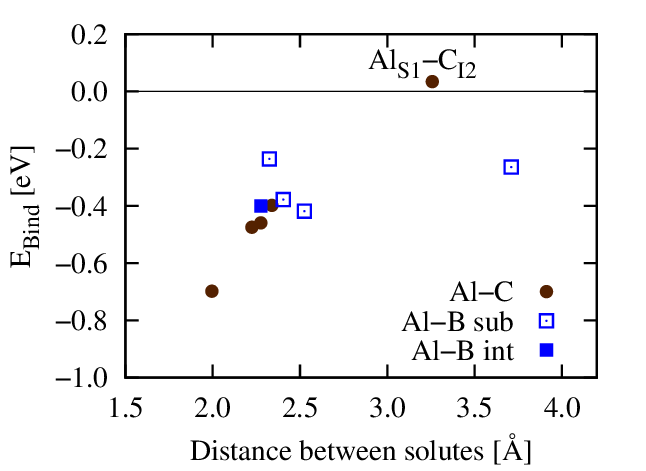}
\caption{}
\end{subfigure}
\begin{subfigure}[c]{1.0\linewidth}
\includegraphics[width=1.0\linewidth]{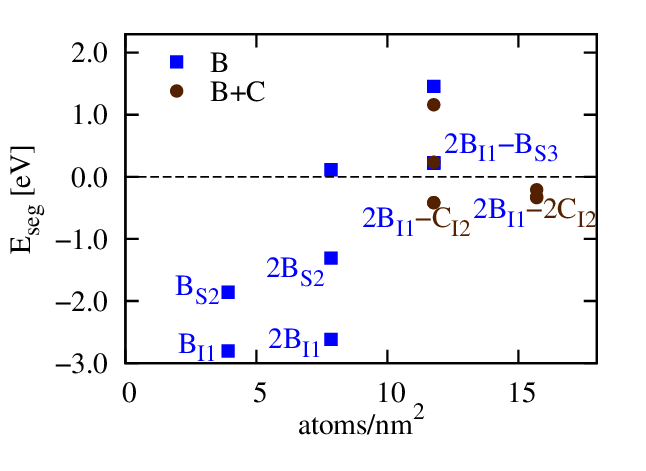}
\caption{}
\end{subfigure}
\caption{\textbf{Solute interactions at GB.} In a), the binding energies of Al with B and with C are shown as a function of the distance between the solutes. Panel b) presents the co-segregation energies of B and C atoms to the GB, where important configurations are named.}
\label{fig:gbsegcov}
\end{figure}

The binding energies of Al in its preferred position and either B or C are shown in Fig. \ref{fig:gbsegcov}~a). For Al and B in their lowest energy sites, S1 and I1, respectively, the binding energy is determined to $-0.40\,\unit{eV}$, which is repulsive. If B is considered in one of the substitutional sites, the binding energy varies between $-0.42\,\unit{eV}$ (Al in S1 and B in S2) and $-0.24\,\unit{eV}$ (Al in S1 and B in S3). For C, multiple repulsive constellations are found with binding energies in the range of $-0.70$ to $-0.40\,\unit{eV}$ for small distances of <2.5~\AA{} between Al and C. For the constellation of Al at site S1 and C at site I2, with a distance of $\sim$3.3~\AA{}, a slightly positive binding energy is obtained. However, this is not the site where C exhibits the strongest segregation tendency. For both C and B interaction with Al, the magnitudes of the binding energies are mostly larger than the segregation tendency of Al to the GB. The strong repulsive interactions of Al with either B or C imply that co-segregation of these elements is highly unlikely. This suggests that Al is repelled from the GB, since both C and B exhibit a much stronger segregation tendency in comparison to Al.

Another effect that comes into play with solute interactions is the coverage dependence, which refers to changes in the segregation energy with increasing coverage at the GB. This may be computed using Eqs.~\ref{eq:eseg} and \ref{eq:esegsubB}, where $E_{gb,i}^x$ denotes the total energy of the structure with N+1 solutes at the GB and $E_{gb}$ the structure with N solutes at the GB. For the lowest solute coverage we take 0.5~ML which corresponds to the results shown in Fig.~\ref{fig:gbseg} and add atoms to preferential segregation sites. The resulting coverage dependent segregation energies are displayed in Fig.~\ref{fig:gbsegcov}~b. As B has by far the strongest segregation energy, the starting structure contains a B atom at the GB. Adding a second B atom leads only to a slightly lower segregation tendency for the interstitial positions (2B$_\mathrm{I1}$), whereas in the substitutional case, the decrease is somewhat stronger (B$_\mathrm{S2}$ to 2B$_\mathrm{S2}$). Adding a third B atom, however, leads to a large decrease in segregation tendency to even positive values (the lowest energy configuration 2B$_\mathrm{I1}$-B$_\mathrm{S3}$ is at +0.22~eV). This means that additional B segregation above a coverage of 8~atoms/nm$^2$ becomes unfavourable. When adding C to the GB to site I2 with already two B atoms present in position I1 gives a negative segregation energy of -0.42~eV. Even for an additional C atom in I2, a negative segregation energy of -0.33~eV is obtained. This indicates that although B shows strong segregation tendencies to different positions at the GB, at 1~ML coverage or about 8~atoms/nm$^2$ a maximum enrichment is reached due to solute interactions.

The enrichment of solutes at GBs in equilibrium can be described with a modified McLean isotherm, but under experimental conditions kinetic processes often influence GB segregation. Therefore, we compute the enrichment of solutes at the GB by considering segregation kinetics via our recently published model \cite{scheiber2018kinetics,scheiber2020solute}. As input parameters we use grains with a radius of $100\,\mu m$ that is further discretized into shells of  $1\,\mu m$ thickness. Although the bi-crystals are in the range of cm, the system size is sufficient because the diffusion length of all involved solutes is below $100\,\mu m$ in the chosen heat treatment. The GB thickness is taken as $0.8\,nm$, which is the region where segregation energies are negative. The diffusion data is given in Table~\ref{tab:diffdata} and for the heat treatment we consider cooling down from $1800~\unit{K}$ with the experimentally informed \cite{scheiber2018kinetics} modified Newton's law of cooling. The cooling rate of $r=0.017 s^{1/2}$ is chosen such that room temperature is reached after 24 hours (see top panel in Fig. \ref{fig:kinseg}) similar to the fabrication of the bi-crystal specimen. The bulk concentration of the solutes was taken from the wet chemical analysis. The segregation energies necessary were taken from the coverage dependence shown in Fig.~\ref{fig:gbsegcov}~b, i.e. for B site I1 is considered and for C the site I2.

\begin{figure}[htbp]
\includegraphics[width=0.98\linewidth]{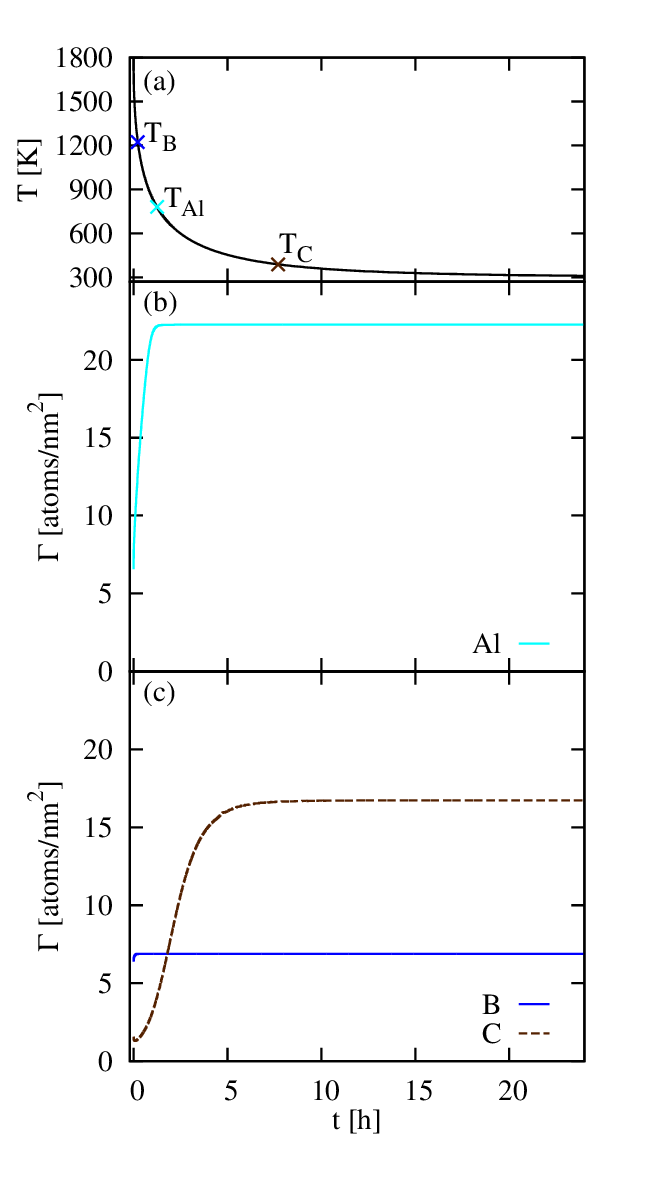}
\caption{\textbf{Computed grain boundary segregation kinetics.} a) Time-temperature curve during cooling with the temperatures $T_{sol}$ marked at which segregation for the solutes is limited kinetically. b) The resulting enrichment of Al atoms without interaction with other solutes. c) Competing temperature dependent segregation of C an B while the bottom panel shows the enrichment for competing B and C segregation.}
\label{fig:kinseg}
\end{figure}

The resulting time and temperature dependent enrichment of solutes at the GB is shown in Fig.~\ref{fig:kinseg}. Due to the strong repulsive interactions of Al with B and C, we investigate Al segregating on its own in the center panel (Fig.~\ref{fig:kinseg}~b)), while we consider mutual segregation for B and C in the bottom panel (Fig.~\ref{fig:kinseg}~c)). The GB enriches with Al for temperatures above $800\,\unit{K}$, but for temperatures below, segregation is kinetically limited. At this stage, the excess concentration of Al is $22\,\unit{atoms/nm^2}$. Here it is important to note that for Al no coverage dependency was considered. For B, the GB enrichment takes place relatively fast at high temperatures because of the strong segregation tendencies and stops at a GB excess of $7\,\unit{atoms/nm^2}$ for temperatures below $1200\,\unit{K}$ due to kinetic reasons. This happens already for higher temperatures than for Al because the concentration in the bulk is much lower than for Al and for GB enrichment, B needs to diffuse from farther away from the GB. Contrary, C is able to enrich until temperatures of only $390\,\unit{K}$ due to faster diffusion and higher content in the bulk than for B so that a final GB excess of $16\,\unit{atoms/nm^2}$ is reached. The fact that B enriches already at higher temperatures than Al reduces the possibility of Al segregation even more because B is already present at the GB in a temperature range where Al segregation takes place.

\begin{table}[htbp]
    \centering
    \begin{tabular}{c|ccc}
        Solute & $D_0$ [m/s$^2$] & $E_a$ [eV] & Ref. \\
        \hline
        Al & $1.80\times 10^{-4}$ & $2.364$ & \cite{mehrer1990diffusion} \\
        C & $3.94\times 10^{-7}$ & $0.831$ & \cite{mehrer1990diffusion} \\
        B & $3.19\times 10^{-7}$ & $2.310$ & \cite{fors_nature_2008}
    \end{tabular}
    \caption{Diffusion data employed for the segregation kinetics simulations.}
    \label{tab:diffdata}
\end{table}

\section*{Discussion}
In this study, the atomic structure, composition and co-segregation mechanisms of Al, C and B at a $\Sigma 5\,(3\,1\,0) [0\,0\,1]$ bcc-Fe GB were investigated by correlating aberration corrected HAADF-STEM observations, APT measurements and first-principles DFT calculations. The observed atomic structure of the $\Sigma 5\,(3\,1\,0) [0\,0\,1]$ GB consists of kite-type structural units and agrees to investigations by Medlin et~al. \cite{medlin_defect_2017}. Globally, the GB appeared straight and close to the symmetric orientation, but atomic resolved observations indicated the presence of defects such as steps or disconnections. Further, in close vicintiy to a GB precipitate steps are incorporated into the GB to compensate the local curvature. Interestingly, this GB segment does not break up into $\{2\,1\,0\}$ and $\{3\,1\,0\}$ facets (as observed by Medlin et~al. \cite{medlin_defect_2017}), but retains the $(3\,1\,0)$ kite-type structural units. The reason is that the inclination angle of only $7^\circ$ from the symmetric orientation seems not to be sufficient to promote faceting.

Our compositional analysis of the GB revealed a depletion of Al at the interface, where locally the Al concentration reduces by $\sim 1.5~at.\%$ at the GB. This is in contradiction to first-principles calculations of Yasua et~al. \cite{yuasa_first-principles_2013} and Geng et~al. \cite{geng_influence_2001} as well as Scheiber et~al. \cite{scheiber2020solute}, who observed the segregation of Al. However, it should be mentioned, that they investigated a pure Fe-Al system. From our APT measurements, we were able to detect the segregation of B and C. Taking these elements into our calculations, we showed that both C and B have much stronger segregation tendencies than Al (Fig.~\ref{fig:gbseg}), so that already simple site competition arguments suggest a depletion of Al due to competition with B for substitutional sites. Explicit calculation of solute interactions revealed strong repulsive interactions between Al-C and Al-B. By considering the kinetics of segregation, it is clear that B is the first solute to segregate to the GB, which prevented Al enrichment and explained the depletion of Al. The composition evaluated with the kinetic model and first principles segregation energies is substantially higher than the composition observed with APT. This is attributed to co-segregation effects between B with itself, C with itself and B with C. In fact, evaluation of these co-segregation effects showed that increased coverage of B decreases the segregation tendency at coverages above 1~ML. It is not possible to implement these interactions into the current kinetic model, still, it can be inferred that co-segregation effects limit enrichment of B and C. The observed enrichment kinetics should hold true nonetheless, i.e. a first fast enrichment of B, followed by slower enrichment of C that continues on until lower temperatures. Even though co-segregation effects are known in general \cite{Erhart1981}, they were not observed in the case of bcc-Fe-Al(C,B) and never explained in a comprehensive manner by combining atomistic simulations and experimental observations. 

Another interesting finding is the observation of locally distorted kite structure. Its origin can be related to the presence of defects, which are aligned along the tilt axis, and to segregation of B to substitutional sites. The assumption that GB defects alter the projected structural units is supported by observations of a periodic distribution of C in our APT analysis. It is known that C preferentially segregates to highly distorted regions such as GB dislocation cores \cite{hristova_solubility_2011}. These defects could also be responsible for the extra contrast observed in the open volume of the kites (indicated by yellow arrows in Fig.\ref{HAADF_FFT}~b)). Hyde et~al. \cite{hyde_atomistic_2005} reported that by the formation of GB dislocation loops a metastable structure having extra Fe atoms in the kite center may appear. 
The distortion of the kites can also have a compositional origin. Our DFT simulations showed that different substitutional sites are also preferred positions for B segregation introducing strong displacements of Fe atoms from their equilibrium positions (Fig.~\ref{fig:Bsub}). STEM image simulations of these structures - especially panel b of Figure~\ref{fig:Bsub} - closely resemble our observations in the red dashed line in Fig.\ref{HAADF_FFT}~a) even so the individual B atoms can not be localized. Furthermore, the formation of Fe dumbbells in the simulated images induced by B segregation could explain the weak contrast in the kite center observed in the experiments (see yellow arrows in Fig.\ref{HAADF_FFT}b). The weaker contrast originates that we do not have a full column of Fe atoms shifted, but partly. Therefore it is highly possible that defects in combination to the B segregation is responsible for the distorted structure of the GB. 

Our results demonstrate that co-segregation effects in the presence of impurities leads to unexpected GB segregation effects and possible GB structural transformations. Considering the interaction of non-metallic elements, here B and C, with substitutional alloying additions, in our case Al, is indispensable to fully capture the entire segregation processes. We discovered that the presence of B and C impurites segregated to the GB is repelling Al from the GB by strong repulsive interactions. Furthermore, the competing mobilities of solutes critically influences the evolution of GB composition. We ultimately show that the presence of non-metallic impurity elements needs to be considered when studying GB segregation effects. However, our work also paves the way for tailoring GB compositing by tuning impurity additions.

\section*{Methods}
\subsection*{Experimental}
For the experiments, bicrystals of an Fe-Al alloy having a $\Sigma 5\,[0\,0\,1]$ GB were grown using an in-house modified Bridgman technique. Samples were extracted from the inital stage of the bicrystal (indicated in supplementary Figure~\ref{supp:EBSD}). Besides the $4\, at.\%$ Al to stabilize the bcc phase \cite{kubaschewski_ironbinary_2013}, wet chemical analysis of the bulk sample revealed $0.001\,at.\%$ B and $0.05\,at.\%$ C.

%\begin{table}[h!]
% \centering
% \begin{adjustbox}{max width=\linewidth}
%\begin{tabular}{|c|c|c|c|c|c|c|c|c|c|}
%\hline
%Element & Al & C & B& O& P& S& Si\\
%\hline
%Content (at.\%) &$4$& $0.05$& $0.001$& $0.22$& $<0.003$& %$0.001$& $0.005$ \\
%\hline
%\end{tabular}
%\end{adjustbox}
%  \caption{Impurity content of the Fe-Al bicrystal.}
%\label{wetchemicanal}
%\end{table}
%

%\begin{figure}[H]
%\centerline{
%\includegraphics[width=8cm]{bi2.png}
%}
%\caption{...caption here...} \label{bicrystal}
%\end{figure} 

The $(0\,0\,1)$ surface of the bicrystal was mechanically polished to obtain a mirror-like area. To visualize the GB, the sample was etched with a solution of $\text{nitric acid}:\text{ethanol} =1:9$ (volume fraction) for $5\,s$. 
The GB showed globally a flat and straight character (Supplementary Figure~\ref{supp:EBSD}~b)). Information about the GB type were acquired form EBSD analysis (Supplementary Figure~\ref{supp:EBSD}~b) and c)), which was carried out by means of a ThermoFisher Scios 2 dual beam FIB/SEM equipped with an EDAX Velocity EBSD camera. 

From the characterized area, several TEM specimens were prepared by FIB. HAADF-STEM images were conducted using a C$_s$ probe-corrected FEI Titan Themis 60-300 operated at $300\,kV$ for imaging and $120\,kV$ for STEM-EDX. Using a semiconvergence angle of $17\,mrad$ and a camera length of $100\,mm$ resulted in a semicollection angle range of $73$ to $352\,mrad$. The probe current was set to $80\,pA$ to prevent beam damages. High resolution images were acquired in a series of at least $30$ frames each recorded with a dwell time of $5\,\mu s$. These frames were stacked by means of cross correlation using the FEI Velox 2.8 software. STEM simulation were performed by the Prismatic Software package \cite{ophus_fast_2017}.
Finally, APT measurements were conducted in a CAMECA LEAP $5000\,XR$ operating in laser-pulsed mode. The pulse rate was $200\unit{kHz}$ with a energy of $30\unit{pJ}$ and the temperature of the specimen was $40\unit{K}$.

\subsection*{Computational Details}
The first principles density functional theory calculations in this study were performed using the Vienna Ab-initio Simulation Package (VASP), which employ projector augmented wave-functions~\cite{kresse1993ab,Kresse1996,Kresse1996a,Kresse1999}. We have chosen the PBE exchange-correlation functional~\cite{perdew1996generalized,perdew1998perdew} as it has been shown that it is suited best to describe Fe~\cite{Haas2009}. The k-point density for all involved simulation cells was set as close as possible to 40~k-points/\AA{} and the energy cutoff to 400~eV. As a result, the lattice parameter of Fe was determined as 2.8386~\AA{}, which is in accordance with DFT literature~\cite{scheiber_ab_2016,jin2014study}. The simulation cells for GBs are build up of two slabs that are misoriented to each other so that they form a GB where they join. On the other side, the slabs are separated by a vacuum layer of at least 8~\AA{} that proved to be large enough to prevent interactions of the two surfaces.

\bibliographystyle{naturemag}
\bibliography{Article}

%During the EBSD data acquisition, the acceleration voltage %was set to $20\,kV$. The working distance was $14\,mm$ and %the frame size for EBSD mapping was set to $10\,\mu m \times %15\,\mu m$ with a step size of $20\,nm$.  
%In~\ref{appendix} the datils are described. 
\section*{Acknowledgement}
The authors gratefully acknowledge the financial support under the scope of the COMET program within the K2 Center “Integrated Computational Material, Process and Product Engineering (IC-MPPE)” (Project No 859480). This program is supported by the Austrian Federal Ministries for Transport, Innovation and Technology (BMVIT) and for Digital and Economic Affairs (BMDW), represented by the Austrian research funding association (FFG), and the federal states of Styria, Upper Austria and Tyrol. G. D. gratefully acknowledges support by the ERC Advanced Grant GB-Correlate (Grant Agreement 787446).

\section*{Author contribution}
G. D. and L. R. designed the project and secured funding. A. A. conducted the electron microscopy experiments including STEM and FIB/SEM.\\ X. Z. conducted the APT experiments. D. S. performed the simulations. C. L., L. R. and B. G. contributed to the data analysis, C. L. and L. R. as well as G. D. revised the paper.

\newpage
\beginsupplement
\section*{Supplementary}\label{supplementary}

\begin{figure*}[htbp]
\includegraphics[width=0.98\linewidth]{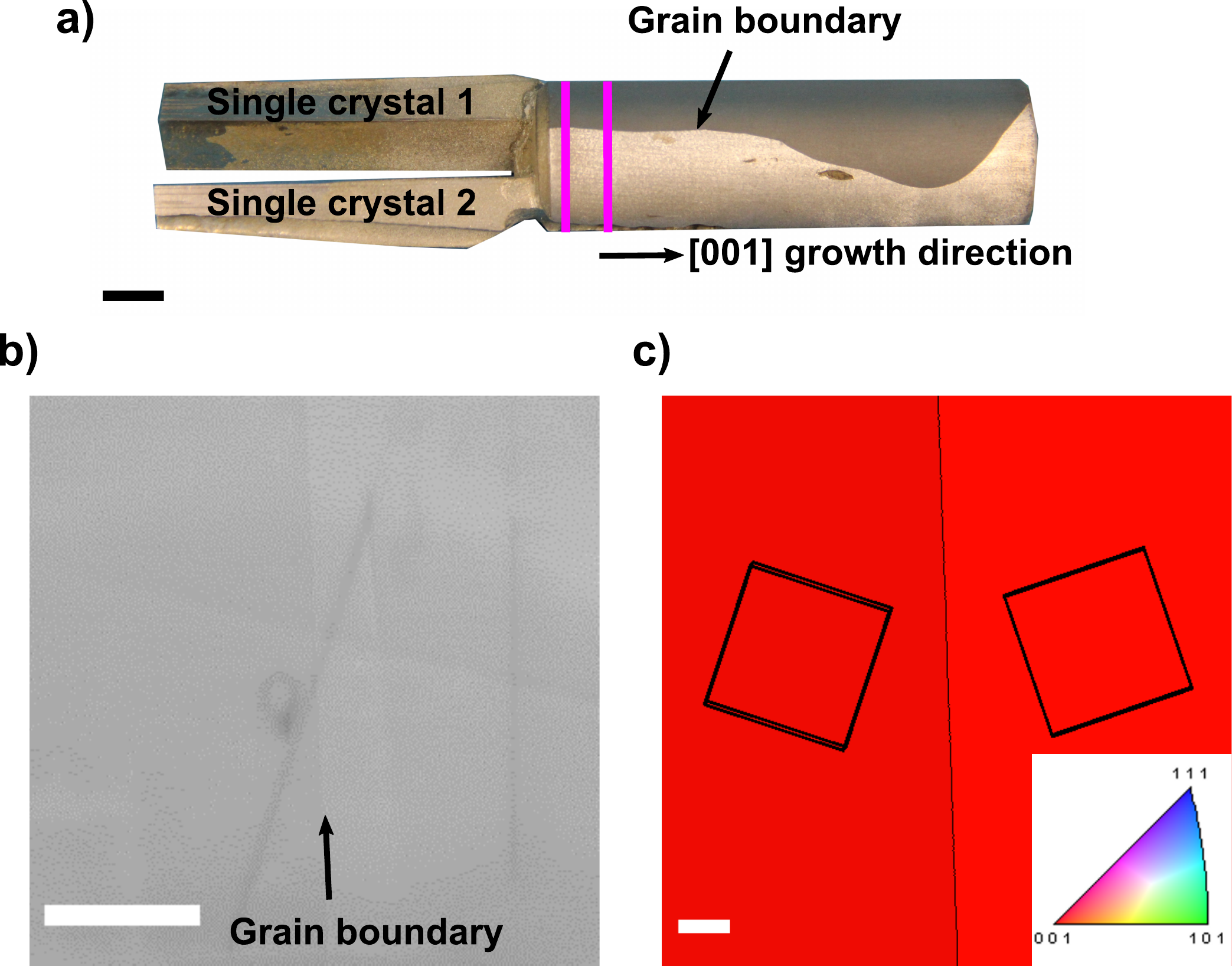}
\caption{\textbf{Global characterization of the GB.} a) Side view image of the bicrystal including the single crystal ingots to fabricate it. The growth direction is $[0\,0\,1]$. Samples were investigated from the initial part (labelled by magenta lines), where the GB started to grew. b) SEM image of the bicrystal top surface (along the $[0\,0\,1]$ tilt axis) after final polishing. The GB is indicated by the black arrow. The GB runs straight and shows no steps or curvatures at the micron scale. EBSD scan of the bicrystal shows a clear $[0\,0\,1]$ texture of both grains. In each grain the orientation of the unit cells is shown by the black rectangles indicating a symmetric misorientation of $~37.5^\circ$. Further analsis of the polefigure the GB plane was obtained to be $(3\,1\,0)$ into both grains. The Scale bar in a) is $10\,m m$ in b) $5\,\mu m$ and in c) $1\,\mu m$.}
\label{supp:EBSD}
\end{figure*}

\begin{figure*}[htbp]
  \includegraphics[width=0.98\linewidth]{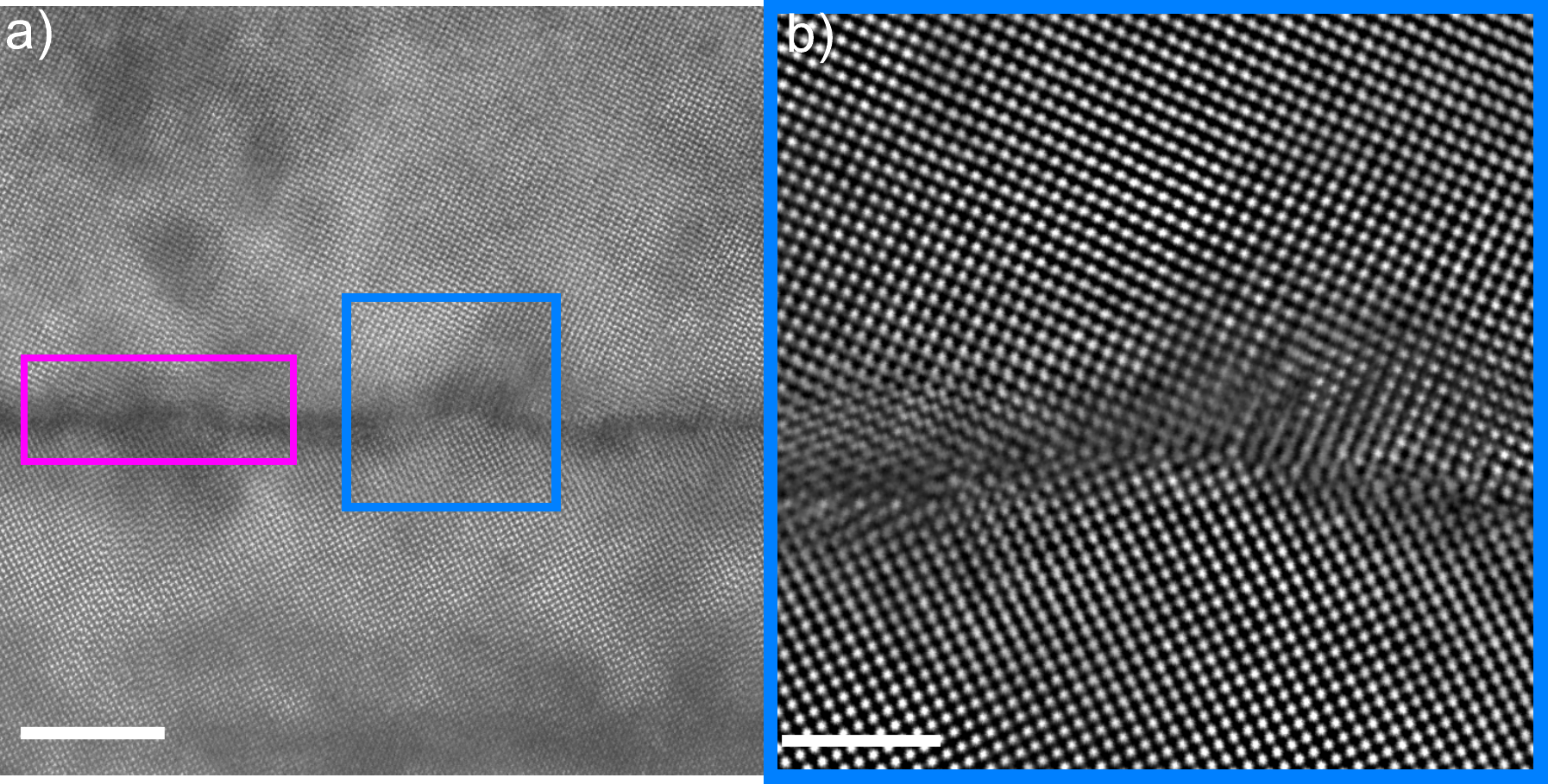}
\caption{\textbf{High resolution imaging of distorted GB structure.} a) HAADF-STEM image of the GB showing small steps (magenta rectangle), where the boundary plane shows facetting. Besides the small steps, large distortions are highlighted in the blue box and a higher magnified image is shown in b). The large step caused a large amount of strain - especially onto the upper grain. Thereby, the kite-structure is not visible anymore. The Scale bar in a) is $5\,nm$ and in b) $2\,nm$.}
\label{supp:HAADF}
\end{figure*}

\begin{figure*}[htbp]
  \includegraphics[width=0.98\linewidth]{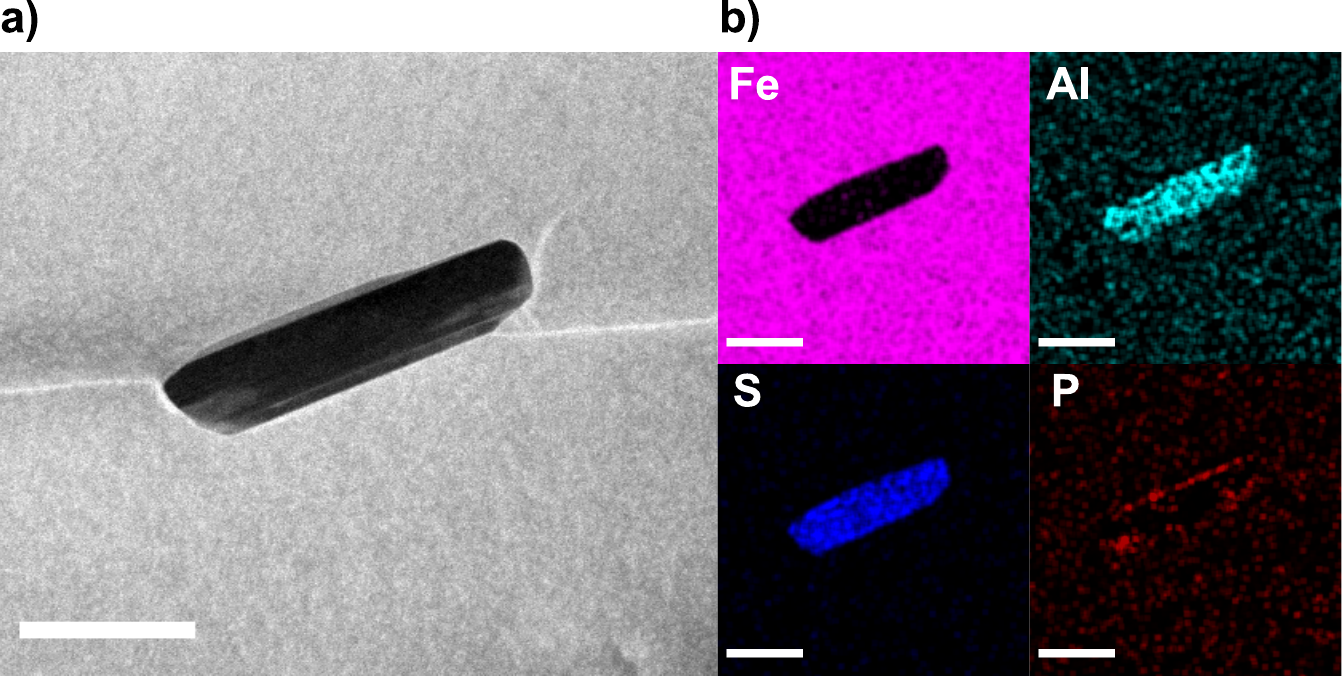}
\caption{\textbf{Formation of precipitates at the GB.} a) HAADF-STEM image of the GB intersecting with a large precipitate. b) The corresponding EDS elemental maps for Fe, Al, S and P.  The Scale bar in a) and b) is $100\,nm$.}
\label{supp:EDS}
\end{figure*}

\end{otherlanguage}
\end{document}